\documentclass[a4paper, 12pt]{article}
\usepackage{subfigure}
\usepackage{wrapfig}
\usepackage[utf8]{inputenc}
\usepackage{amsmath}
\usepackage{amsfonts}
\usepackage{mathtools}
\usepackage{bbm}
\usepackage{chngcntr}
\usepackage{amsmath}
\usepackage{amsfonts}
\usepackage{amssymb}
\usepackage{mathrsfs}
\usepackage{bm}
\usepackage{amsthm}
\usepackage{textcomp}
\usepackage[utf8]{inputenc}
\usepackage{eurosym}
\usepackage{scalerel,stackengine}
\stackMath
\newcommand\reallywidehat[1]{%
\savestack{\tmpbox}{\stretchto{%
  \scaleto{%
    \scalerel*[\widthof{\ensuremath{#1}}]{\kern-.6pt\bigwedge\kern-.6pt}%
    {\rule[-\textheight/2]{1ex}{\textheight}}
  }{\textheight}%
}{0.5ex}}%
\stackon[1pt]{#1}{\tmpbox}%
}

\newtheorem{theorem}{Theorem}
\newtheorem{proposition}{Proposition}[theorem]

\newtheorem{definition}{Definition}[theorem]
\newtheorem{assumption}{Assumption}
\newtheorem{lemma}{Lemma}[theorem]

\counterwithin{theorem}{section}
\counterwithin{definition}{section}
\counterwithin{lemma}{section}
\counterwithin{fact}{section}
\counterwithin{proposition}{section}
\counterwithin{corollary}{section}
\counterwithin{assumption}{section}
\DeclareMathOperator*{\argmin}{argmin} 
 
\usepackage{amssymb}
\usepackage{graphicx}
\usepackage{parallel}
\usepackage{lipsum}
\usepackage{multicol}
\usepackage[paperwidth=8.5in, paperheight=11in,top=2.5cm,bottom=2.5cm,left=2.5cm,right=2.5cm]{geometry}
\usepackage{multirow}
\usepackage[table]{xcolor} 
\usepackage{amsmath}
\usepackage{amsfonts}
\usepackage{amssymb}
\usepackage{mathrsfs}
\usepackage{multicol}
\usepackage{bm}
\usepackage{booktabs}
\usepackage[hidelinks,colorlinks=true,linkcolor=blue,citecolor=blue]{hyperref}

\usepackage{natbib}
\usepackage{listings}
\usepackage{color} 
\definecolor{mygreen}{RGB}{28,172,0} 
\definecolor{mylilas}{RGB}{170,55,241}
\usepackage{pdflscape}
\usepackage{lscape} 
\usepackage{caption}
\usepackage[title]{appendix}
\usepackage{setspace, caption}
\usepackage{lipsum}
\doublespace   

\title{Pricing Carbon Allowance Options on Futures:\\ 
Insights from High-Frequency Data}

\author{
Simone~Serafini\thanks{Department of Mathematics, University of Bologna, Piazza di Porta San Donato 5, 40126 Bologna, Italy. Email address: simone.serafini@unibo.it} \and Giacomo~Bormetti\thanks{Department of Economics and Management, University of Pavia, Via San Felice al Monastero 5, 27100, Pavia, Italy. Email address: giacomo.bormetti@unipv.it}
} 

\date{}

\begin{document}

\maketitle

\begin{abstract}
\noindent Leveraging a unique dataset of carbon futures option prices traded on the ICE market from December 2015 until December 2020, we present the results from an unprecedented calibration exercise. Within a multifactor stochastic volatility framework with jumps, we employ a three-dimensional pricing kernel compensating for equity and variance components' risk to derive an analytically tractable and numerically practical approach to pricing. To the best of our knowledge, we are the first to provide an estimate of the equity and variance risk premia for the carbon futures option market. We gain insights into daily option and futures dynamics by exploiting the information from tick-by-tick futures trade data. Decomposing the realized measure of futures volatility into continuous and jump components, we employ them as auxiliary variables for estimating futures dynamics via indirect inference. Our approach provides a realistic description of carbon futures price, volatility, and jump dynamics and an insightful understanding of the carbon option market.
\end{abstract}



\section{Introduction}
Over the past few decades, the global reliance on fossil fuels and petrochemical products has drastically increased, accelerating the concentration of greenhouse gases (GHGs) in the atmosphere. This has led to a significant shift in weather patterns, melting ice caps, rising sea levels, and a higher frequency of extreme weather events such as hurricanes, droughts, and floods. The impact of climate change on both the environment and human society has become undeniable, prompting international action to mitigate these effects. One of the key responses has been the establishment of carbon markets, designed to reduce emissions by allowing countries and companies to trade emission permits, primarily for carbon dioxide ($\text{CO}_2$).
The Kyoto Protocol, adopted in 1997 and enforced in 2005, was a groundbreaking initiative aimed at reducing GHG emissions on a global scale. It introduced legally binding targets for industrialized nations to cut their emissions, laying the foundation for carbon trading schemes such as the European Union Emissions Trading Scheme (EU ETS). Since its inception, the EU ETS has played a pivotal role in the global carbon market, evolving into a crucial mechanism for controlling emissions. The EU ETS is the European Commission's primary mechanism for reducing GHG emissions, encompassing large facilities within GHG-intensive industries across the EU. Under the EU ETS, firms must hold enough emission permits by year-end to cover their $\text{CO}_2$ emissions from the previous year. The scheme allows the trading of these permits, influencing $\text{CO}_2$ allowance prices. As countries continue to develop and refine these carbon markets, the economic implications of carbon pricing on energy and commodity markets become increasingly significant, further highlighting the need for collaborative international efforts to tackle climate change.
 Each European Union Allowance (EUA) permits the emission of one tonne of $\text{CO}_2$ or an equivalent amount of another GHG. Operating as a cap-and-trade system, the EU ETS ensures that marginal abatement costs are equalized among firms by allowing the trading of allowances under a predetermined emissions cap. Firms exceeding their allocated allowances can either purchase additional allowances from the market or implement emission reduction measures. Conversely, surplus allowances can be sold, making the right to emit $\text{CO}_2$ a tradable asset. The cap-and-trade approach, therefore, provides both flexibility and economic efficiency in achieving emission reduction targets.
The EU ETS was established in 2005 under EU Directive 2003/87/EC and has undergone several phases. 
In addition to the spot market for these certificates, a substantial market exists for EUA derivatives, such as futures and options, primarily traded on the European Energy Exchange (EEX) and the Intercontinental Exchange (ICE). Futures contracts, particularly those close to expiration, exhibit high liquidity and trading volumes, making them a focal point for both market participants and researchers.
Numerous studies have investigated the EU ETS, focusing on its market characteristics and the behaviour of EUA prices. Early research examined price determinants, revealing correlations with energy prices, climatic factors, and economic events (e.g., \cite{Mansanet}, \cite{Alberola2}, \cite{Alberola3}, \cite{CHEVALLIERRiskFactors}, \cite{HINTERMANN201043}, \cite{Hammoudeh}). For instance, \cite{HINTERMANN201043} detailed the mechanism by which the cap-and-trade system equalizes marginal abatement costs, while \cite{Alberola2} and \cite{CHEVALLIERRiskFactors} analyzed the influence of energy prices and weather conditions on EUA prices.
EUA price volatility has attracted considerable interest, with studies like \cite{BenthCarbon} and \cite{KIM2017714} using stochastic volatility models to analyze the dynamics of EUA futures.  Additionally, research on the valuation of carbon futures options have been conducted, starting by \cite{Carmona2011RiskNeutralMF}, who proposed reduced-form models for risk-neutral allowance price dynamics. More recently, \cite{Yang} highlighted the importance of accounting for price jumps in option valuation, while \cite{Fang} introduced a mixture lognormal price approach for valuation.
Despite these advances, few studies have leveraged intraday data for analyzing the EU ETS. One of the earliest investigations, by \cite{ChevallierRV}, employed realized volatility measures to capture long-memory effects in EUA futures, demonstrating that HAR models outperform traditional GARCH models. Subsequent research by \cite{WaldeMarPriceFormationHF} and \cite{HITZEMANN_intradayVol} focused on price formation and intraday volatility, finding partial evidence of classical U-shaped intraday patterns and significant volatility responses to announcements. \cite{BenschCabrera} further demonstrated the superior performance of HAR models in forecasting realized volatility using high-frequency data.

Our study contributes to this literature in key aspects, such as modeling EUA futures realized volatility (RV) and pricing carbon futures options with a pricing kernel compensating equity and variance components' risks.  Using high-frequency data from ICE, we construct and analyze the RV of carbon futures, decomposing it into continuous and discontinuous components and identifying the number and amplitude of price jumps occurring on a given day. We develop discrete-time models for the dynamics of the continuous and jump RV components that we will later use as auxiliary processes for statistical inference. For option pricing, due to the limited liquidity of the market, we propose the following framework. We specify a multifactor stochastic volatility model under the historical probability $\mathbb{P}$, estimated via the indirect inference method (\cite{Gourioeux1993}) using the HAR class models on RV as auxiliary models. A unique aspect of our work is the introduction of a new pricing kernel with three risk premia, one associated with the equity risk and the remaining two compensating for the risk of the variance components. We derive the mapping of the model parameters under the historical measure to a risk-neutral counterpart following no arbitrage considerations. The model's analytical tractability allows the implementation of numerically practical fast Fourier pricing techniques, such as the SINC method of \cite{SINCBasch}, for option valuation and calibration. We present the results from a pricing exercise based on historical data for the options on futures market during the Phase 3 from the ICE market. To our knowledge, we are the first to provide a numerical estimate of the equity and variance risk premia for the carbon allowance market.\\
The rest of the paper proceeds as follows: Section~\ref{sec:Mod_frm} details the modeling framework, including the stochastic volatility dynamics and the transition to the risk-neutral measure through the three-dimensional pricing kernel. Section~\ref{sec::Empirical App} presents the empirical analysis, covering realized volatility construction, option pricing, and risk-premia calibration. We compare our approach to the state-of-the-art discrete-time pricing model, LHARG-ARJ, by \cite{Alitabetal}. Finally, Section~\ref{sec:CH2_conclusion} discusses the key findings.

\section{Modeling Framework}\label{sec:Mod_frm}

This section outlines the model used for pricing carbon futures options. We begin by specifying the dynamics of the affine multi-component stochastic volatility model with jumps (2-SVJ) under the historical measure $\mathbb{P}$. Then, we map the parameters of this class of models into the risk-neutral counterpart through a three-dimensional pricing kernel compensating equity and variance components' risks. We derive the analytical expression of the model characteristic function (CF) under the pricing measure $\mathbb{Q}$.

\subsection{Futures dynamics under $\mathbb{P}$}
We assume that the log-price of the carbon futures under the historical probability $\mathbb{P}$ follows the dynamics 
\begin{equation}\label{eq::phisical_SDE}
\begin{aligned}
d X_t=~&  \mu(\sigma_{1,t},\sigma_{2,t})dt  +\sigma_{1, t} d W_{1,t}^{X}+\sigma_{2, t} d W_{2, t}^{X}+c_X d J_t \\
d \sigma_{1, t}^2=~& m_1(\sigma_{1,t})dt + \Lambda_1\sigma_{1,t} dW^{\sigma_1}_{1,t}\\
d \sigma_{2, t}^2=~& m_2(\sigma_{2,t})dt + \Lambda_2\sigma_{2,t} dW^{\sigma_2}_{2,t}
\end{aligned}
\end{equation}
where the Brownian motions are correlated in the following way 
\begin{equation*}
    \begin{aligned}
        &\text{corr}(d W_{1,t}^{X}, dW^{\sigma_1}_{1,t}) = \rho_1, \\
        &\text{corr}(d W_{2,t}^{X}, dW^{\sigma_2}_{2,t}) = \rho_2, \\
    \end{aligned}
\end{equation*}
all other cross-correlations are zero. The drift $\mu(\sigma_{1,t},\sigma_{2,t})$ takes the usual form of a constant term minus the convexity correction from the logarithmic price transform. $J$ is an independent Poisson process with intensity $\lambda$ and jump size $c_X \sim \mathcal{N}(\mu_J, \sigma_J^2)$. We denote for simplicity
\begin{equation*}
    \begin{aligned}
        m_1(\sigma_1) = & ~\kappa_1\left(\omega_1-\sigma_{1, t}^2\right), \\
        m_2(\sigma_2) = &~ \kappa_2\left(\omega_2-\sigma_{2, t}^2\right),
    \end{aligned}
\end{equation*}
where $k_j$ is the speed of mean reversion and $\omega_j$ is the long-run level of the variance components for $j=1,2$. This model yields several well-known specifications used in option pricing.
\begin{itemize}
\item With $\sigma_{2,t}=0$ and $J = 0$, we retrieve the classic stochastic volatility (SV) \cite{Heston1993} model.
\item With $\sigma_{2,t}=0$ and jumps, we have the stochastic volatility with price jump (SVJ) model used in \cite{Bates} and \cite{Bakshi}.
\item With $\sigma_{2,t}>0$ and $J^{X} = 0$, we obtain the double Heston (2-SV) model used by \cite{Christoffen2009}.
\item With $\sigma_{2,t}>0$ and allowing jumps, we derive the 2-SVJ model developed in \cite{Bates2000}.
\end{itemize}

\noindent The model in Equation~(\ref{eq::phisical_SDE}) belongs to the class of affine processes, along with the different specifications. Affine processes are popular in option pricing because their CF is available in closed form. This allows for the use of numerically practical Fourier pricing techniques, such as the SINC method of \cite{SINCBasch}, enabling fast and accurate option pricing. Estimating this class of continuous time models can be challenging. One typically needs to resort to filtering techniques combined with numerically cumbersome estimation procedures. As we will later detail, we leverage the information from high-frequency data to design an estimation approach based on indirect inference rendering the procedure viable and practical. 

\subsection{Dynamics under risk-neutral probability $\mathbb{Q}$}\label{sec::riskneutral}

We now analyze the risk-neutral dynamics associated with the model in Equation~(\ref{eq::phisical_SDE}). To maintain analytical tractability, we risk-neutralize the model using a pricing kernel from the exponential affine family, which offers the flexibility to incorporate multiple risk premia compensating price and variance factors' risks. We focus on the 2-SVJ model, the most general case. The risk-neutral versions of other models can be derived by setting the relevant parameters to zero.
\begin{assumption}
    We assume that the dynamics under $\mathbb{P}$ of the ECF log-price $X_t$ follows the 2-SVJ dynamic specified in Equation~(\ref{eq::phisical_SDE}). The pricing kernel $M_t$ takes the form \begin{equation}\label{PricingKernel}
M_t=M_0\left(\frac{p_t}{p_0}\right)^\phi e^{\int_0^t \delta\left(\sigma_{1,s}, \sigma_{2,s}\right) ds+\psi_1(\sigma_{1,t}^2-\sigma_{1,0}^2) + \psi_2(\sigma_{2,t}^2 - \sigma_{2,0}^2) },
\end{equation}
where $\phi$ is the parameter controlling the aversion of price risk while $\psi_1$ and $\psi_2$ are the variance risk premia for the components $\sigma_{1,t}$ and $\sigma_{2,t}$, respectively. The function $\delta(\cdot)$ controls time preferences. Here, the pricing kernel is written as a function of the observed price,  $p_t = e^{X_t}$.
\end{assumption}
\noindent The main motivation for incorporating two factors for the variance is based on empirical evidence.  Later, we will show that the best models estimated using the indirect inference approach have two volatility factors.  A single factor model for the variance  fails to replicate the statistical features captured by the HAR models (for the empirical evidence, please refer to the Supplemental Information (SI)).
A similar pricing kernel, but with a single volatility factor, is introduced by \cite{ChristoffenPricing} and used in \cite{BANDIReno}. This pricing kernel is monotonically decreasing in prices when $\phi < 0$ (assuming the variance constant) and monotonically increasing in both variance factors when $\psi_1>0$ and $\psi_2 > 0$ (keeping prices constant). Since typically high prices are associated with periods of high variance, when projected on prices the pricing kernel may reproduce the empirically observed U-shaped behavior. The non-monotonicity of the pricing kernel is able to reconcile a variety of empirical facts and puzzles. The following proposition provides a closed-form representation of the risk-neutral dynamics and a characterization of the mapping from the historical to the risk-neutral parameters in terms of the equity and variance components' risk premia.
\begin{proposition}\label{Prop::pricing_Kernel}
Assume that we have price dynamics under $\mathbb{P}$ specified in Equation~(\ref{eq::phisical_SDE}) and a pricing kernel as specified in Equation~(\ref{PricingKernel}). Denoting by $r$ the risk-free rate, the risk-neutral dynamics under $\mathbb{Q}$ is given by
\begin{equation}\label{Risk_neutralSDE}
\begin{aligned}
&d X_t=  {\left(r-\lambda^*\mathbb{E}[e^{c_X^*}-1]-\frac{1}{2}\left(\sigma_{1, t}^2+\sigma_{2, t}^2\right)\right) d t } +\sigma_{1, t}^2 d \Tilde{W}_{1,t}^{X}+\sigma_{2, t} d \Tilde{W}_{2, t}^{X}+c_X^* d \tilde{J}_t \\
&d \sigma_{1, t}^2= m_1^*(\sigma_{1,t})dt + \Lambda_1\sigma_{1,t} d\tilde{W}^{\sigma_1}_{1,t}\\
&d \sigma_{2, t}^2= m_2^*(\sigma_{2,t})dt + \Lambda_2\sigma_{2,t} d\tilde{W}^{\sigma_2}_{2,t}
\end{aligned}
\end{equation}
where we have correlated Brownian motions
\begin{equation*}
    \begin{aligned}
        &\text{corr}(d \Tilde{W}_{1,t}^{X}, d\tilde{W}^{\sigma_1}_{1,t}) = \rho_1, \\
        &\text{corr}(d \Tilde{W}_{2,t}^{X}, d\tilde{W}^{\sigma_2}_{2,t}) = \rho_2, \\
    \end{aligned}
\end{equation*}
while $\tilde{J}_t$ is an independent Poisson process with risk-neutral intensity $\lambda^*$ and risk-neutral jump size $c_X^*$. The following results hold
\begin{equation*}
    \begin{aligned}
        & m_1^{*}(\sigma_{1,t}) - m_1(\sigma_{1,t}) = \sigma_{1,t}^2(\phi \rho_1 \Lambda_1  + \psi_1 \Lambda_1^2) \\
        & m_2^{*}(\sigma_{2,t}) - m_2(\sigma_{2,t}) = \sigma_{2,t}^2( \phi \rho_2 \Lambda_2  + \psi_2 \Lambda_2^2) \\
        & \lambda^{*} = \lambda \mathbb{E}[e^{\phi c_X}]
        \end{aligned}
\end{equation*}
and, for all $u \in \mathbb{R}$ 
\begin{equation*}
    \mathbb{E}[e^{iuc^*_X}] = \frac{\mathbb{E}\left[e^{iuc_X} (e^{\phi c_X})\right]}{\mathbb{E}\left[ e^{\phi c_X }\right]} =\exp{\left((\mu_J+\phi \sigma^2_J) i u - \frac{1}{2}\sigma^2_J u^2 \right)}\,.
\end{equation*}
Then, we have $c^*_X \sim N(\mu_J + \phi \sigma^2_J, \sigma^2_J)$.
\end{proposition}
\begin{proof} See Appendix~\ref{app:proofs}.
\end{proof}

We conclude this section by reporting the closed-form expression of the CF for the risk-neutral 2-SVJ model.
\begin{lemma}\label{lem::chf}
    Under the model specified by the dynamics in Equation~(\ref{Risk_neutralSDE}), the time $t$ conditional log-return CF $\hat{f}_x (z, x_t, \sigma_{1,t}^2, \sigma_{2,t}^2, t, T) = \mathbb{E}^{\mathbb{Q}}[e^{izX_T} | \mathcal{F}_t]$, for $T > t$, is given by
\begin{equation*}
\begin{aligned}
& \log \hat{f}_x(z ; \tau)=i\left(x_t+(r-q) \tau\right)z \\
& \quad+\sum_{j=1,2}\left(A_j^x(z ; \tau)+B_j^x(z ; \tau) \sigma_{j, t}^2\right)+C_{J}^x(z ; \tau),
\end{aligned}
\end{equation*}
where $\tau = T-t$, $z \in \mathbb{C}$ and the coefficients $A_j^x$, $B_j^x$ and $C_J^x$ are specified in the following way
\begin{equation*}
\begin{aligned}
& A_j^x(z ; \tau)=\frac{\kappa_j \omega_j}{\Lambda_j^2}\left[\left(c_j-d_j\right) \tau-2 \log \left(\frac{1-g_j e^{-d_j \tau}}{1-g_j}\right)\right] \\
& B_j^x(z, \tau)=\frac{c_j-d_j}{\Lambda_j^2} \frac{1-e^{-d_j \tau}}{1-g_j e^{-d_j \tau}} \\
& C_{J}^x(z ; \tau)=\lambda \tau\left(\theta^{J}(z )-1-i \mathbb{E}[e^{c_X^*}-1] z\right)
\end{aligned}
\end{equation*}
where $\theta^{J}(z )$ is the CF of the jump size. We have defined the parameters
\begin{equation*}
\begin{aligned}
c_j &= \kappa_j - i z \rho_j \Lambda_j, \\
d_j &= \sqrt{c_j^2 + z (i + z) \Lambda_j^2}, \\
g_j &= \frac{c_j - d_j}{c_j + d_j}.
\end{aligned}
\end{equation*}
\end{lemma}
\begin{proof}
    See Appendix~\ref{app:proofs}.
\end{proof}

\section{Empirical Analyses}\label{sec::Empirical App}
\subsection{Data}

\begin{figure}[h!]
    \centering
    \includegraphics[width = \textwidth]{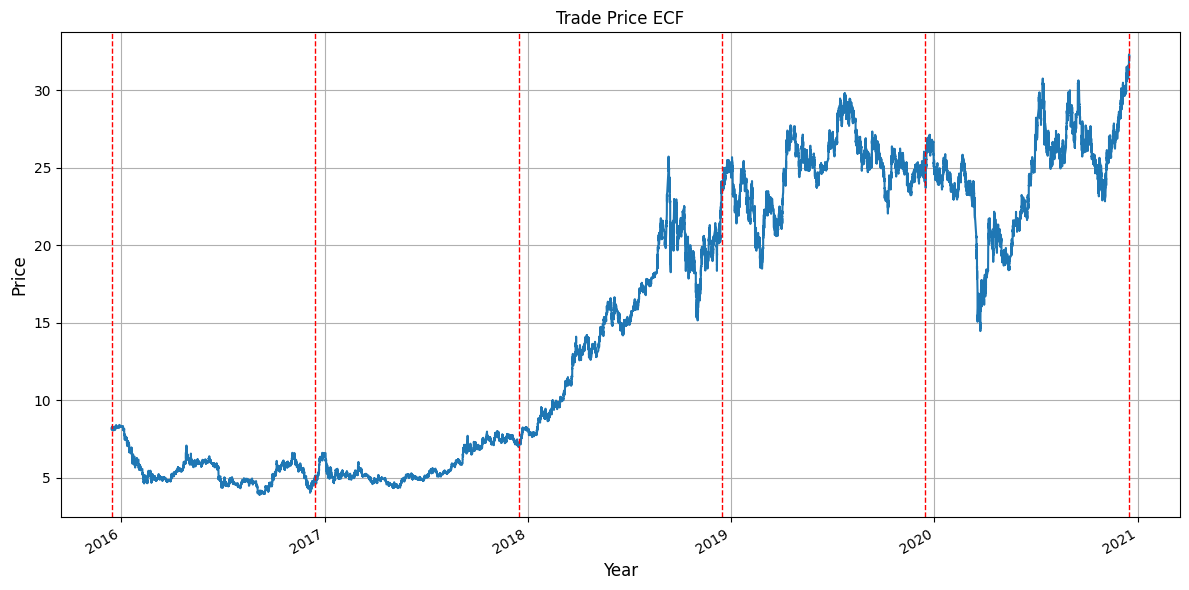}
    \caption{ ECF price; red lines indicate the roll-overs to the next maturity for futures contracts.}
    \label{fig::EUA Prices}
\end{figure}

\begin{figure}[h!]
    \centering
    \includegraphics[width = \textwidth]{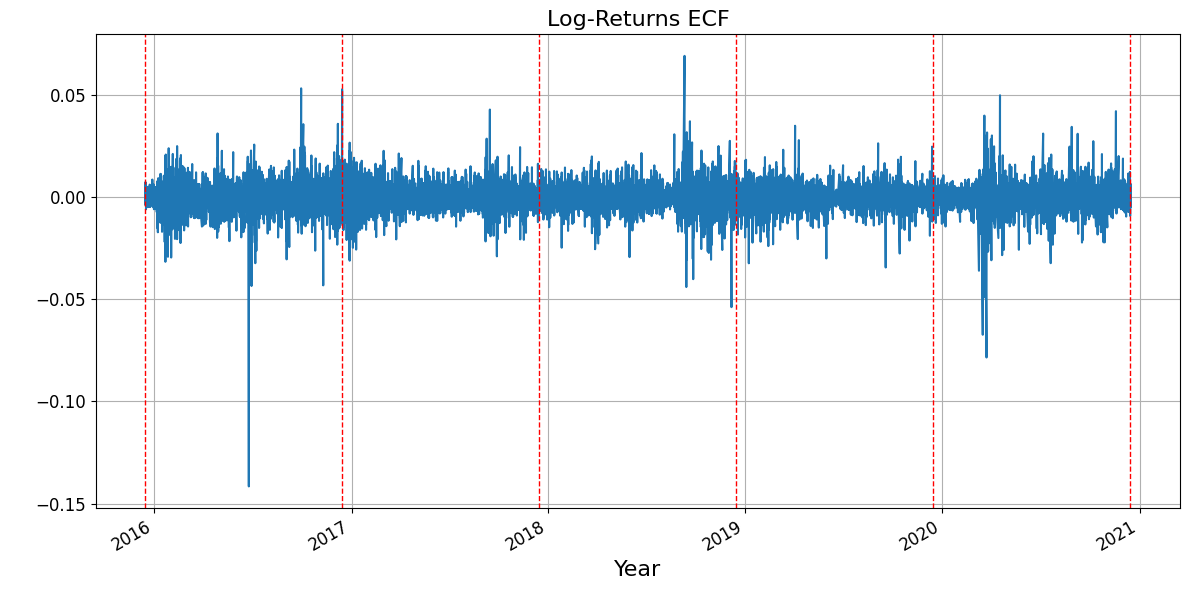}
    \caption{ECF log-returns; red lines indicate the roll-overs to the next maturity for futures contracts.}
    \label{fig:: Returns}
\end{figure}

\begin{figure}[h!]
    \centering
    \includegraphics[width = \textwidth]{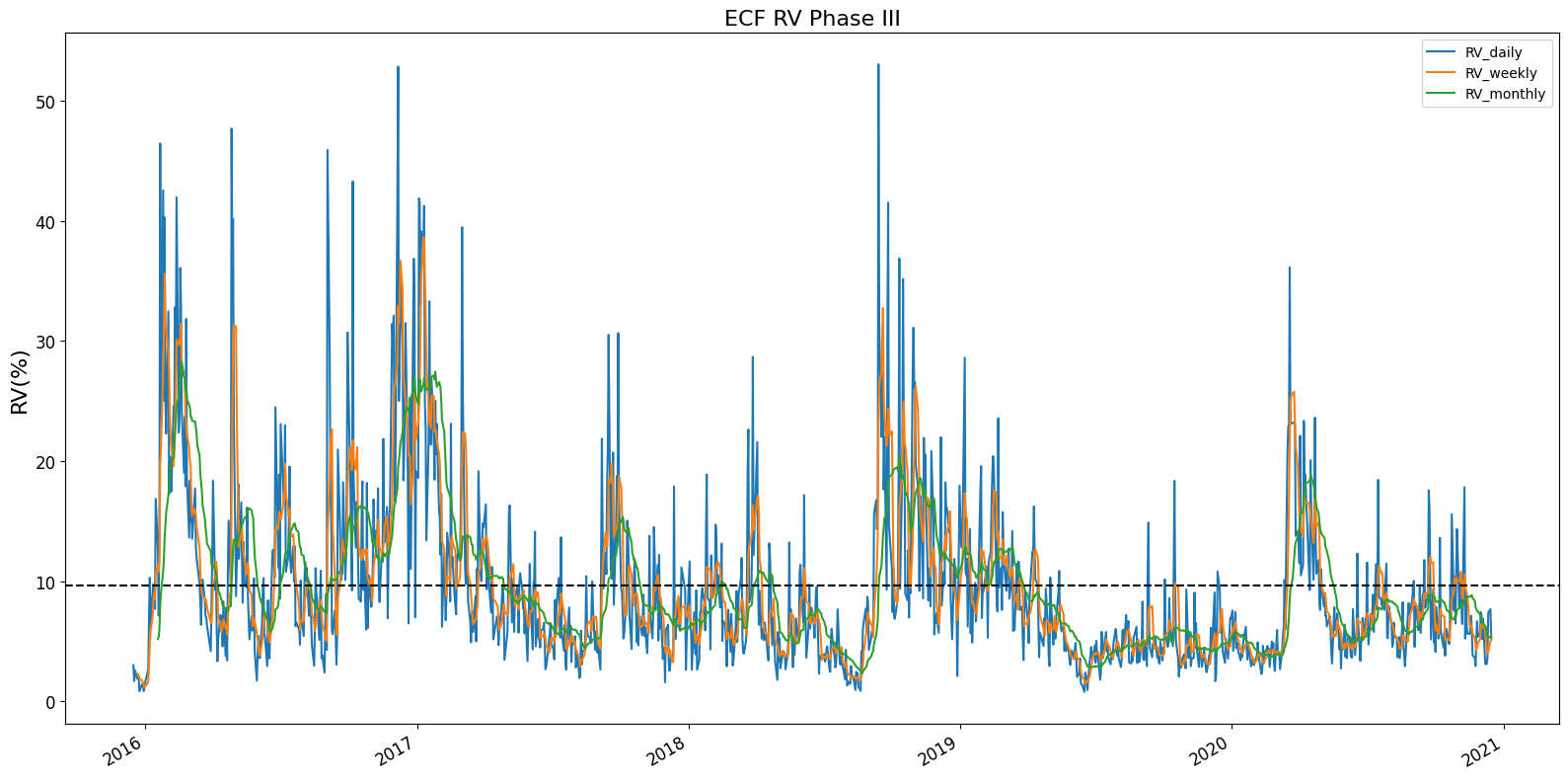}
    \caption{RV daily, weekly and monthly of the ECF futures calculated from five-minute log-returns. The horizontal black line indicates the mean value.}
    \label{fig::RVdwm}
\end{figure}

For empirical analysis, we used tick-by-tick data on futures and options on futures trade prices from ICE covering the period of Phase 3 of the EU ETS: from December 15, 2015, to December 15, 2020. The reference ticker for the futures instrument is ECF, followed by a month and year abbreviation. We focus exclusively on December-expiry futures, designated as ECFZ, since they are the most liquid in this market. Moreover, the options on futures traded on the exchange have only the December futures as underlying. In order to construct the time series of trade prices, we roll-over to the next December futures the day before the expiry date. From the tick-by-tick data, we construct the time series of futures prices and log-returns at five-minute frequency. Given the low market liquidity on certain days (particularly in the early years of our sample and when futures are far from maturity), we opted to use a five-minute frequency in order to get rid of the possible effects of microstructure noise when constructing realized volatility measures. 
We present descriptive statistics of ECF prices and intraday returns in Table~(\ref{tab:descriptive_stats}). The market appears volatile, exhibiting a high standard deviation in log returns, with prices varying from under 5 euros to over 30 euros over the past six years. The high kurtosis in log returns also emphasizes the extreme events that occurred during this period.
Trading occurs from 7:00 AM to 5:00 PM GMT, resulting in 120 five-minute intervals per day. The dataset comprises 155,238 observations spanning 1,283 trading days, including futures contracts on six underlying futures, from ECFZ15 to ECFZ21. Figure~(\ref{fig::EUA Prices}) shows the ECF prices during Phase 3, with vertical red lines marking the rollovers to the next December futures contracts. We observe a number of extreme negative returns that align with significant geopolitical events concerning the EU, particularly Brexit in 2016, which raised concerns about the UK’s potential exit from the EU ETS, see \cite{EUETSBrexit} for details. The volatile conditions of 2020, driven by the COVID-19 pandemic and widespread lockdown announcements across Europe, created significant fluctuations in energy markets and a sharp decline in prices, impacting the emissions market due to its correlation with other commodities, as outlined by \cite{COVIDEUETS}. A similar scenario unfolded in 2022 with the Russia-Ukraine war, which also resulted in pronounced volatility and instability in this market since its connection to the gas market, as highlighted in \cite{cornago2022eu}.
Figure~(\ref{fig:: Returns}) illustrates the log-returns, highlighting volatility clustering and heteroskedasticity. 
\begin{table}[h!]
\centering
\begin{tabular}{lcc}
\toprule
 & Futures prices & Log-returns(\%) \\
\midrule
Mean & 15.16 & 0.00 \\
Std & 8.96 & 0.28 \\
Minimum & 3.88 & -14.16 \\
Maximum & 32.33 & 6.89 \\
Skewness & 0.09 & -0.95 \\
Kurtosis & 1.33 & 64.87 \\
\bottomrule
\end{tabular}\caption{Descriptive statistics of futures prices and log-returns (in percentage) at five-minute sampling frequency.}\label{tab:descriptive_stats}
\end{table}
\noindent Figure~(\ref{fig::RVdwm}) illustrates the RV aggregated components -- at daily, weekly, and monthly level following~\cite{CorsiRV} -- in daily percentage unit, after the data pre-treatment outlined in the SI). Weekly and monthly RVs are smoothed versions of the daily realized variance. The figure highlights volatility clustering, low speed of relaxation after abrupt variance increases, and significant spikes. Higher volatility periods are often triggered by major geopolitical events and influenced by auction effects. The mean realized variance is 9.6\%, corresponding to an annualized mean volatility of 49.5\% (in line with values in Table~(\ref{tab:descriptive_stats})). 
The SI offers additional exploratory analyses of the dataset, covering liquidity, plot of the estimated quadratic variation components, and the leverage effect.
\noindent During the Phase 3 of the EUA market, options are traded under the ticker O:ECF with quarterly maturities in March, June, September, and December. December options expire one trading day before the underlying futures. Our dataset includes 520 options traded from December 2016 to December 2020, comprising 263 puts and 257 calls. We define moneyness as $m = K/p_t$. The mean moneyness for calls is 1.24, while for puts it is 0.76, indicating that, on average, the options are out-of-the-money (OTM). Evidence shows that many deep-OTM options were traded during turbulent market conditions, such as the 2020 price drop due to COVID lockdowns, for both hedging and speculative purposes.

\subsection{Model estimation}

For estimation, we follow the route traced in~\cite{CorsiRenoJBES} and rooted on the indirect inference method. Indirect inference is a simulation-based method for estimating the parameters of a structural model, consisting of two stages. First, an auxiliary model is fitted to the observed data. Next, a binding function maps the structural model parameters to the auxiliary ones. The method iteratively simulates the structural model by varying its parameter values in order  to minimize the distance between the auxiliary models parameter estimated on the synthetic data and those obtained from the historical time series (more details in the SI). In our setting, leveraging high-frequency data, we construct realized volatility measures, separating them into continuous and jump components. Additionally, using the method described by \cite{Andersen2010}, we estimate the number of intraday jumps and the associated sizes. Following \cite{CorsiRenoJBES}, we employ the LHAR-CJ class of models as auxiliary models to estimate the structural multifactor stochastic volatility model under the real-world probability.

The following subsections are structured as follows. We begin with the results of the discrete-time analysis of realized volatility using the LHAR-CJ models, detailed in the SI, assessing both in-sample and out-of-sample performance. This step leads us to the selection of the optimal model for describing the volatility dynamics in the futures market. Estimates from the LHAR-CJ models are then used within the indirect inference framework to estimate the model specified in Equation~(\ref{eq::phisical_SDE}).
\subsubsection{Auxiliary model performance on market data}\label{subsec:lharcjres}
The estimation process of the LHAR-CJ model and nested specifications is straightforward and performed through ordinary least square (OLS) with Newey-West covariance corrections.
We pre-process the data following the methodology reviewed in the SI. Table~(\ref{tab:model_estimation}) shows that the HAR model yields daily and weekly coefficients of similar magnitude, while the monthly component is much smaller and not significant. The models are estimated in the daily percentage convention for all the variables involved. The LHAR model shows that the coefficients for daily and weekly negative returns are the most significant, despite having lower $t$-statistics than the volatility coefficients, with the weekly coefficient being less significant. This may be due to a small correlation between lagged $RV_t$ values and negative returns, as illustrated in the SI.
\begin{table}[h!]
\centering
\begin{tabular}{lccc}
\toprule
 & HAR & LHAR & LHAR-CJ \\
\midrule
$c$ & 0.229 & 0.237 & 0.284 \\
    & (4.889) & (4.676) & (5.449) \\
$\beta^{(d)}$ & 0.422 & 0.391 & 0.379 \\
    & (12.977) & (11.931)  & (14.182) \\
 $\beta^{(w)}$ &  0.422 & 0.384 & 0.258 \\
               & (7.685) & (7.686) & (6.064) \\
$\beta^{(m)}$ & 0.043 & 0.076 & -0.008 \\
            & (0.782) & (1.315) & (-0.131) \\
$\gamma^{(d)}$ &  & -0.024 & -0.025 \\
               &  & (-2.895) & (-2.960) \\
$\gamma^{(w)}$ &  & -0.034 & -0.044 \\
               &  & (-2.502) & (-3.343) \\
$\gamma^{(m)}$ &  & -0.002   & -0.018 \\
               &  & (-0.074) & (-0.514) \\
 $\alpha^{(d)}$ &  &  & 0.035\\
 &  &  & (1.679) \\
$\alpha^{(w)}$ &  &  & 0.021 \\
 &  &  & (0.869) \\
$\alpha^{(m)}$ &  &  & 0.144 \\
 &  &  & (5.381) \\
\bottomrule
\end{tabular}\caption{In-sample estimates of the LHAR-CJ model and nested models. Parentheses contain the $t$-statistics. }\label{tab:model_estimation}
\end{table}
\noindent This suggests a weak leverage effect. We fitted all models with both positive and negative leverage, since the commodity market may exhibit an 'inverse leverage' effect, where volatility correlates more with positive returns (see, for example, \cite{CARNERO_leverage}). Our analysis indicates that the negative leverage effect is slightly more significant than the positive one; therefore, we opted to use negative returns as covariates in the LHAR-CJ class models, for details, see the SI. In the complete LHAR-CJ model, we find that the volatility and leverage coefficients exhibit similar characteristics, while the jump component has statistical significant monthly and daily coefficients. This could be because the market experiences many phases during which the jump component remains high and persistent.
We evaluate the in-sample performance of the model using the following metrics: AIC, BIC and adjusted R-squared. Table~(\ref{tab:models_in_sample}) shows that while the LHAR and LHAR-CJ have the same AIC, the BIC metrics select the HAR and LHAR specifications. Conversely, the LHAR-CJ has a slightly higher R-squared value. We compared the HAR class models to benchmark models, including AR and ARFIMA, which are commonly used to analyze long-memory properties of volatility in this market, as done in previous studies such as by \cite{BenschCabrera}. Notably, the heterogeneous aggregation significantly improves the performance, with the HAR model class outperforming the others.
\begin{table}[h!]
\centering
\begin{tabular}{lccc}
\toprule
Model   & AIC & BIC      &  $R^2_{\text{adj}}$ \\
\midrule
HAR  & 1423 & 1443&0.605 \\
LHAR    & \textbf{1407} & \textbf{1443} & 0.618 \\
LHAR-CJ     & \textbf{1407} & 1459 & \textbf{0.622} \\
AR(22)  & 1478 & 1602&0.460 \\
ARFIMA  & 1476 & 1601&0.448 \\
\bottomrule
\end{tabular}
\caption{Full in-sample model comparison.}
\label{tab:models_in_sample}
\end{table}
As a final check to support our selection of the auxiliary model in indirect inference, we evaluate the model's out-of-sample forecasting ability, by focusing on forecasting the tomorrow realized variance, $\hat{V}_{t+1}$. We perform one-day ahead forecasts, re-estimating the model daily at time $t$ using an expanding window of observations up to $t-1$. To assess the out-of-sample performance, we forecast realized volatility from December 1, 2019, to December 15, 2020, using a test set of 264 days. We measure forecasting accuracy with root mean squared error (RMSE), mean absolute error (MAE), the QLIKE metric introduced by \cite{PattonQlike}, and the $\text{R}^2$ from Mincer-Zarnowicz regressions.
Table~(\ref{tab:modelsoof}) shows that the LHAR and LHAR-CJ models outperform the classic HAR, while the HAR class models exceed the performance of simpler AR and ARFIMA models, as before.
\begin{table}[h!]
\centering
\begin{tabular}{lcccc}
\toprule
Model   & MSE      & MAE & $R^2$ & QLIKE \\
\midrule
HAR  & 0.127 & 0.285 & 0.540  & 1.606\\
LHAR    & \textbf{0.122} & \textbf{0.280} & 0.553 & 1.604\\
LHAR-CJ     & \textbf{0.122} & \textbf{0.280} & \textbf{0.559} & \textbf{1.603}\\
AR(22) & 0.150&0.308 &0.478 & 1.607\\
ARFIMA &0.135 &0.296 & 0.519& 1.606\\
\bottomrule
\end{tabular}
\caption{Out-of-sample results.}
\label{tab:modelsoof}
\end{table}

\subsubsection{Estimation via indirect inference }\label{sec::IFest}
This section presents the results of the estimation via indirect inference of the multifactor stochastic volatility model with jumps. We confirm, as noted in \cite{CorsiRenoJBES} for the S\&P500, that the simplest HAR model is not suitable for calibrating one-factor stochastic volatility models (SV and SVJ) in the EU-ETS market as well. Detailed results are provided in the SI. We therefore focus on the stochastic volatility model with two factors (2-SV and 2-SVJ). We begin our analysis by calibrating a 2-SV model specifying the dynamics in Equation~(\ref{eq::phisical_SDE}) with J = 0 and without leverage. We employ the HAR model as the auxiliary model. For identification reasons, we set $\omega_1 = \omega_2 = \omega$ and $\mu(\sigma_{1,t}, \sigma_{2,t}) = 0$. The former restriction is dictated by the fact that the auxiliary model is sensitive solely to the long-run level of the realized volatility and cannot disentangle the long-run level of the single variance components. The latter assumption is consistent with the fact that, asymptotically, the contribution of the drift component to the quadratic variation is negligible. For the estimation of the structural model with no jump components via the auxiliary HAR/LHAR models, we employ as obervables the time series of $\hat{C}_t$, the empirical estimator of the continuous component of the quadratic variation. 
The structural model parameters are scaled and presented in daily non-percentage terms. For ease of comparison, we  follow the same convention adopted for the LHARG-ARJ model~\cite{Alitabetal}, that we will use as benchmark model in the calibration exercise. The implied parameters follows the convention of Section~\ref{subsec:lharcjres}. Errors in the estimates are computed using the asymptotic result from \cite{Gour1996}.
We use variance targeting techniques to reduce the parameter space dimension. We define each $\omega$ as half the mean of the continuous component of the quadratic variation, while for one vol-of-vol parameter, we target it based on the unconditional variance of the volatility dynamics 
\begin{equation*}
    \Lambda_2 = \sqrt{\kappa_2\left(2  \frac{\emph{Var}(\hat{C})}{\omega} - \frac{\Lambda_1}{\kappa_1}\right)},
\end{equation*}
Therefore, we need to estimate three parameters. Table~(\ref{tab::HARIF}) shows the estimates and the corresponding implied HAR coefficients.
\begin{table}[h!]
\setlength{\tabcolsep}{3em}
\centering

\begin{tabular}{lll}
\toprule
\multicolumn{3}{c}{\textbf{Structural model:}} \\
\midrule
\multicolumn{3}{c}{
    $\begin{aligned}
d X_t=~&   \sigma_{1, t} d W_{1,t}^{X}+\sigma_{2, t} d W_{2, t}^{X} \\
d \sigma_{1, t}^2=~& m_1(\sigma_{1,t})dt + \Lambda_1\sigma_{1,t} dW^{\sigma_1}_{1,t}\\
d \sigma_{2, t}^2=~& m_2(\sigma_{2,t})dt + \Lambda_2\sigma_{2,t} dW^{\sigma_2}_{2,t}
\end{aligned}$
} \\
\midrule
Parameter & Estimates & Std. Errors \\
\midrule
$\kappa_1$ & 5.14e-02&5e-04\\
$\kappa_2$ & 2.63 & 3e-02 \\
$[\omega]$ & 4.32e-04 & \\
$\Lambda_1$ & 9.72e-03 & 5e-05 \\
$[\Lambda_2]$ & 3.29e-02 & \\
$\chi^2$ & 1.21e-03& \\
\midrule
\multicolumn{3}{c}{\textbf{Auxiliary model:}} \\
\midrule
\multicolumn{3}{c}{
    $\log \widehat{\mathrm{C}}_{t+h}^{(h)}=c  +\beta^{(d)} \log \widehat{\mathrm{C}}_t+\beta^{(w)} \log \widehat{\mathrm{C}}_t^{(5)}+\beta^{(m)} \log \widehat{\mathrm{C}}_t^{(22)} + \varepsilon_t^{(h)}$
} \\
\midrule
Parameter & Estimated & Implied \\
\midrule
$c$           & 0.245 & 0.258 \\
$\beta^{(d)}$ & 0.465 & 0.456 \\
$\beta^{(w)}$ & 0.377 & 0.368 \\
$\beta^{(m)}$ & 0.029 & 0.043 \\
$\sigma_\varepsilon^2$ & 0.204 & 0.178 \\
\bottomrule
\end{tabular}\caption{Structural (2-SV) and auxiliary (HAR) model estimation results. Parameters between squared parentheses are fixed by targeting. \label{tab::HARIF}}
\end{table}
The two-factor model effectively reproduces the HAR coefficients, all estimates are inside the standard deviation of the implied parameters. Notably, we observe a fast mean-reverting factor associated with high vol-of-vol, alongside a slower mean-reverting factor with a half-life of approximately 20 days. This combination produces the desired volatility persistence. These findings qualitatively align with those in \cite{CorsiRenoJBES} and \cite{RossiDemag} for the S\&P 500, where the HAR model has a very strong convergence and the implied parameters are well reproduced by a 2-SV model without leverage. However, our analysis quantitatively shows a higher mean-reverting factor and larger vol-of-vol, consistently with the high volatility of ECF market with respect to the S\&P 500.\\

\noindent We then estimate the full 2-SVJ model. We incorporate a nonzero correlation coefficient to introduce a leverage effect by utilizing negative returns in the auxiliary model.
For the jump component, we can target the intensity, the mean and standard deviation of jumps due to the fact that the number of intraday jumps and their size are made observable quantities by the procedure outlined by~\cite{Andersen2007}. This enables us to use the LHAR model as an auxiliary model instead of the LHAR-CJ model, which requires much larger computational time and many simulations in the indirect inference method to have enough statistics for fitting the jump parameters, as noted by \cite{RossiDemag}. 
Furthermore, since targeting allows us to reduce the dimensionality of the optimisation problem, we decrease the number of auxiliary parameters retaining only the most significant components of the LHAR model -- the daily and the weekly-- while maintaning the identificability of the structural parameters. This leads to a reduction in the variability of the estimation. The implied parameters are well reproduced.
 We report our estimate in Table~(\ref{tab:2SVJ-IF}), where we note the magnitudes of the mean-reversion and vol-of-vol coefficients remain unchanged with respect to the 2-SV case, and we observe two negative leverage components. The largest leverage component corresponds to the slow mean-reversion factor, while the smallest is linked to the fast factor. This likely reflects an overshooting effect, as noted by \cite{CorsiRenoJBES}, where negative correlations at low frequencies are often low or positive, and vice versa.
 
\begin{table}[h!]
\setlength{\tabcolsep}{3em}
\begin{center}
\begin{tabular}{lll}
\toprule
\multicolumn{3}{c}{\textbf{Structural model:}} \\
\midrule
\multicolumn{3}{c}{
    $\begin{aligned}
    dX_t =~& \sigma_{1,t} dW_t^1 + \sigma_{2,t} dW_t^2 + dN_t\\
    d \sigma_{1, t}^2=~& m_1(\sigma_{1,t})dt + \Lambda_1\sigma_{1,t} dW^{\sigma_1}_{1,t}\\
d \sigma_{2, t}^2=~& m_2(\sigma_{2,t})dt + \Lambda_2\sigma_{2,t} dW^{\sigma_2}_{2,t}
    \end{aligned}$
} \\
\midrule
Parameter & Estimates & Std.Errors\\
\midrule
$\kappa_1$ & { }3.93e-02& 1e-03 \\
$\kappa_2$ & { }2.03 & 2e-02\\
$[\omega]$ & { }4.31e-04 & \\
$\Lambda_1$ & { }8.28e-03 & 1e-04 \\
$[\Lambda_2]$ & { }3.20e-02 & \\
$\rho_1$ & -0.82 & 7e-03\\
$\rho_2$ & -0.11 & 5e-03\\
$[\lambda_J]$ & { }0.72 & \\
$[\mu_J]$ & -7.9e-03 & \\
$[\sigma_J]$ & { }8.5e-03 & \\
$\chi^2$ & { }1.43e-04 & \\
\midrule
\multicolumn{3}{c}{\textbf{Auxiliary model:}} \\
\midrule
\multicolumn{3}{c}{  $\begin{aligned}
\log \widehat{\mathrm{C}}_{t+h}^{(h)}=c & +\beta^{(d)} \log \widehat{\mathrm{C}}_t+\beta^{(w)} \log \widehat{\mathrm{C}}_t^{(5)}\\
& +\gamma^{(d)} r_t^{-}+\gamma^{(w)} r_t^{(5)-}+\varepsilon_t^{(h)}
\end{aligned}$
} \\
\midrule
Parameter & Estimated & Implied \\
\midrule
$c$ & { }0.279 & { }0.278 \\
$\beta^{(d)}$ &  { }0.429 & { }0.424 \\
$\beta^{(w)}$ &  { }0.386 & { }0.381 \\
$\gamma^{(d)}$ & -0.026 &-0.018 \\
$\gamma^{(w)}$ & -0.040 & -0.036 \\
$\sigma_\varepsilon^2$ &{ }0.199 & { }0.199 \\
\bottomrule
\end{tabular}\caption{Structural (2-SVJ) and auxiliary (LHAR) model estimation results. Parameters between squared parentheses are fixed by targeting.}
\label{tab:2SVJ-IF}
\end{center}
\end{table}

\subsection{Option Pricing}
In this section, we present the option pricing exercise based on the 2-SVJ model. We compare the pricing performance with the LHARG-ARJ model introduced in~\cite{Alitabetal}. The latter is a discrete-time model of the futures log-prices with observable volatility and jumps. It belongs to the class of RV heterogeneous auto-regressive gamma  processes~\cite{CorsiFusVE,Maje2015} extended to include a jump component with time-varying
intensity. A flexible specification of the pricing kernel compensates for equity, volatility,
and jump risks. Then, it provides a state-of-the-art benchmark model that, as well as our approach, leverages tick-by-tick data in the construction of the realized volatility measures.  
We also analyze the performance of the 2-SV model, hence not considering the jump component. We apply a standard filter to our sample that excludes options with maturities shorter than 1 day or longer than 365 days. We define the implied volatility of the option on futures data in agreement with the market practice.
\begin{definition}
    The implied volatility, $IV$, of an allowance futures option is the volatility that equates the option's market price under the \cite{black76} model. This model assumes that the dynamic of the underlying futures price $F(t,T_F)$, where $T_F$ is the futures settlement date, follows a geometric Brownian motion. The risk-neutral price of the European allowances futures call and put option with strike $K$ and maturity $T_O\in (t, T) $ is given by 
    \begin{equation*}
\begin{aligned}
& C = e^{-r \tau} \left( F(t,T_F) \cdot N(d_1) - K \cdot N(d_2) \right)
\\
& P = e^{-r\tau} \left( K \cdot N(-d_2) - F(t,T_F) \cdot N(-d_1) \right),
\end{aligned}
\end{equation*}
where $\tau = T_O - t$ is the option time-to-expiration, $N(\cdot)$ is the cumulative distribution function of the standard normal distribution, with $d_1 =  \frac{\ln\left(\frac{F(t,T_F)}{K}\right) + \frac{1}{2} \sigma^2  \tau}{\sigma \sqrt{\tau}}$ and $d_2 = d_1 - \sigma \sqrt{\tau}$.
\end{definition}

\noindent The $\mathrm{IV}$ in our sample has an average value of 60\%, primarily due to the prevalence of out-of-the-money options.
The mapping to a risk-neutral setting of the 2-SVJ (or 2-SV) model involves three risk premium parameters $(\phi, \psi_1, \psi_2)$, specified in the pricing kernel in Equation~(\ref{PricingKernel}). The arbitrage condition outlined in the Proposition~(\ref{Prop::pricing_Kernel}) defines the condition of these parameters. For the calibration procedure, we adopt a method based on the unconditional minimization of the distance between the market-implied and the model-implied volatility. We calculate option prices and implied volatility associated using the SINC method, leveraging the knowledge of the CF in closed form. Following~\cite{Fang}, we set the risk-free rate and the cost-of-carry to zero.
We obtain the optimal risk premia through the following minimization
\begin{equation*}
        \argmin\limits_{(\phi, \psi_1, \psi_2)}\left\{f_{\text{obj}}((\phi, \psi_1, \psi_2))\right\}\,.
\end{equation*}
The objective function $f_{\text{obj}}$ corresponds to the $\mathbb{L}^2$ norm of the difference between the model implied volatility  and the implied volatility of the market for each option $j$, with $j=1,\ldots, N_{\text{opt}}$. It can be expressed as 
\begin{equation*}
        f_{\text{obj}}((\phi, \psi_1, \psi_2))=\sqrt{\sum\limits_{j=1}^{N_{\text{opt}}}\left(\mathrm{IV}_{j}^{\text{mod}}\left((\phi, \psi_1, \psi_2)\right)-\mathrm{IV}_{j}^{\text{mkt}}\right)^2},
    \end{equation*}
In summary, the numerical procedure for calibration is as follows: We estimate the 2-SVJ model through the indirect inference method, with parameters reported in Section~\ref{sec::IFest}. The mapping to the risk-neutral setting is performed using Proposition~\ref{Prop::pricing_Kernel}, and the model is calibrated by computing prices with the SINC method. We set the model variance to the state of the RV of each specific day during calibration. Conditioning improves the flexibility of the model and its ability to adapt to the changing market conditions.
    
Table~(\ref{tab:unified_params}) reports the calibrated risk premia for the 2-SVJ and 2-SV model. From Proposition~\ref{Prop::pricing_Kernel}, the risk-neutral variance components follow the equation
\begin{equation*}
    d\sigma^2_{j,t} = \kappa_j^*(\omega_j^* - \sigma^2_j) + \Lambda_j\sigma_{j,t}d\tilde{W}^{\sigma_j}_{j,t}\,,
\end{equation*}
for $j = 1, 2$ and 
\begin{equation*}
    \begin{aligned}
        \kappa_j^* &= \kappa_j - \phi \rho_j \Lambda_j  - \psi_j \Lambda_j^2\,,\\
        \omega_j^* &= \omega_j \frac{\kappa_j}{\kappa_j - \phi \rho_j \Lambda_j  - \psi_j \Lambda_j^2}\,. 
    \end{aligned}
\end{equation*}
The latter equations clarify how the risk premia modify the speed of mean reversion and the long run level of each volatility factor under the pricing measure.
Notably, the two premia $\psi_1$ and $\psi_2$ exhibit distinct magnitudes: $\psi_1$ compensates for the risk associated with the variation of $\sigma_{1,t}^2$, which features slow mean reversion and low vol-of-vol. On the other hand, $\psi_2$ compensates the variation of the factor $\sigma_{2,t}^2$, characterized by high mean reversion and high vol-of-vol.
Since, as expected, the equity premium coefficient, $\phi$, is negative and the correlations $\rho_1$ and $\rho_2$ are negative too, risk-neutralization leads to a decrease in the mean reversion speed and an increase in the long-term level of each variance factor. The different amplitude of the variance risk premia then allows to distinguish the long-term risk-neutral level of each factor, differently from what happens under the historical measure where both components contribute equally to the variance long-term mean. The objective function $f_{\text{obj}}$ is of the same order of magnitude in both models, with the 2-SVJ model performing better. Incorporating jumps in the stochastic volatility calibration slightly improves pricing performance.
\begin{table}[h!]
\centering
\begin{tabular}{lcc}
\toprule
\textbf{} & \textbf{2-SV} & \textbf{2-SVJ} \\
\midrule
$\phi$   & -7.46e-03 & -7.35e-03 \\
$\psi_1$ &  2.74e-03 &  2.81e-03 \\
$\psi_2$ &  1.28e-02 &  1.50e-02 \\
$f_{\text{obj}}$ & 4.06 & 4.05 \\
\bottomrule
\end{tabular}
\caption{Calibrated risk premia for the 2-SV and 2-SVJ models.}
\label{tab:unified_params}
\end{table}

\subsection{Pricing performances}
    \label{sec:Option_performance}
    In this section we compare the pricing performance of our approach with the discrete-time LHARG-ARJ model of \cite{Alitabetal}. It is worth stressing that the LHARG-ARJ model is a benchmark in this paper. Nevertheless, it is the first time it has been calibrated and tested on options on carbon futures. From this perspective, our exercise provides a comparative assessment of the pricing performance of two entirely different approaches that both leverage the flexibility of a multi-dimensional specification of the pricing kernel. At variance with the 2-SVJ model, the pricing kernel for the LHARG-ARJ depends parametrically on four risk coefficients. They compensates directional (equity) and non-directional (variance) risks associated to the continuous and discontinuous components of the efficient price process. While compliance with the absence of arbitrage principle constraints the premia corresponding to the directional risks, the variance risk premia, $\nu_c$ and $\nu_j$, will be optimized during the calibration exercise.\\
    The pricing performance is evaluated with the percentage Implied Volatility Root Mean Square Error ($RMSE_{IV}$) introduced by~\cite{Ren97}, computed as
    \begin{equation*}
        RMSE_{IV} = \sqrt{\frac{1}{N_{\text{opt}}}\sum_{j=1}^{N_{\text{opt}}} \left(\mathrm{IV}_{j}^{\text{mod}}-\mathrm{IV}_{j}^{\text{mkt}}\right)^2} \times 100\,,
    \end{equation*}
    where $N_{\text{opt}}$ is the number of options and the other terms represent the market and model implied volatility. The $\mathrm{IV}^{\text{mod}}$ is calculated as outlined in the previous section.
    We evaluate the models across different moneyness $m$ intervals to ensure a comparable number of contracts in each of them. Due to the predominance of OTM and deep-OTM options in our sample, we differentiate between three intervals: the first includes primarily OTM and deep-OTM put options with $m < 0.85$; the second comprises mainly OTM and deep-OTM call options with $m > 1.1$; and the third interval, $0.85 \leq m \leq 1$, features a mix of call and put options with less extreme moneyness.
    In the Appendix~\ref{app:LHARGARJ}, we provide an overview of the estimated parameters and the calibrated risk premia of the LHARG-ARJ model, adding a significant and unprecedented contribution to our analysis.
    
Table~(\ref{tab:performancepricing1}) presents the $RMSE_{IV}$ values for various moneyness levels across the LHARG-ARJ, 2-SV, and 2-SVJ models. All models exhibit similar performance with comparable values across moneyness intervals. Still, the LHARG-ARJ performs slightly better in all cases, especially in the moneyness interval of OTM and deep OTM calls for $m > 1.1$. The interval with the highest errors is the one with moneyness $m<0.85$, consisting of OTM and deep-OTM puts.
\begin{table}[h!]
    \begin{center}
     
        \begin{tabular}{lcccc}
            
            & & \multicolumn{3}{c}{$RMSE_{IV}$ } \\ 
            \cline{3-5}
            Model $\backslash$ Moneyness & & $ m < 0.85$ & $0.85 \leq m \leq 1.1$ & $m > 1.1$ \\ 
            \hline 
            \rowcolor{gray!30} LHARG-ARJ & & 19.25 & 11.17 & 10.12 \\
            2-SVJ & & 19.80 & 11.96 & 11.33 \\
            \rowcolor{gray!30} 2-SV & & 19.98  & 11.95 & 11.47 \\
            \hline
            
        \end{tabular}

    \end{center}
    
    \vspace{3mm}
    
    \caption{ Pricing performances of the LHARG-ARJ, 2-SVJ, and 2-SV models for different moneyness intervals.}\label{tab:performancepricing1}
\end{table}

\noindent We will now assess the performance of the LHARG-ARJ and 2-SVJ models by considering the maturity of options across four intervals: short-term options with $\tau<50$, medium-term options with $50<\tau \leq 90$, longer-term options with $90<\tau \leq 160$, and finally, options with long maturity where $\tau>160$. We report the results in Table~(\ref{tab:performance_moneyness_period}).

The models exhibit similar performance with some differences across maturities. For very short maturities, the LHARG-ARJ model consistently outperforms the 2-SVJ model in all defined moneyness intervals. Both models, as before, consistently exhibit higher $RMSE_{IV}$ for the interval $m<0.85$ across all maturity intervals, likely due to the elevated implied volatility of these put options, which exceeds the sample average. For medium-term options, the 2-SVJ model performs better for options with less extreme moneyness ($0.85 \leq m \leq 1.1)$, while the LHARG-ARJ has lower $RMSE_{IV}$ value in the other two intervals. Notably, in the third interval, the 2-SVJ model only outperforms the LHARG-ARJ model for $m<0.85$ and for the last interval composed of long-term maturities, the LHARG-ARJ model achieves the best results. The LHARG-ARJ model's enhanced performance can be attributed to its flexibility. The autoregressive structure of the centrality coefficient in the gamma specification, makes it highly sensitive and reactive to the state of the market on each day and previous month RV and returns time series. Moreover, allowing the jumps' intensity to have an autoregressive structure improves the fit in extreme moneyness regions. 
\begin{table}[h!]                    
    \begin{center}
                        
    \begin{small} 
                            
                            \begin{tabular}{lrrrrrrr}
                                
                                & \multicolumn{7}{c}{Maturity} \\ \cline{2-8}
                                
                                &  &  &  &  &  &  &  \\
                                
                                Moneyness & \multicolumn{1}{c}{$\tau \leq 50 $ } &  & \multicolumn{1}{c}{$ 50 < \tau \leq 90$} &  & \multicolumn{1}{c}{ $ 90  < \tau \leq 160 $} &  & \multicolumn{1}{c}{$ 160 < \tau$} \\ 
                                
                                \hline 
                                
                                & && && && \\ 
                                
                                Panel A & \multicolumn{7}{c}{LHARG - ARJ Implied Volatility RMSE} \\ 
                                
                                & && && && \\
                                
                                \rowcolor{gray!30} $0.85 \leq m \leq1.1$ &10.50&& 13.64&& 12.09&& 6.40 \\
                                
                                 $m<0.85$ &29.35&& 18.47&& 11.67&& 19.41 \\
                                
                                \rowcolor{gray!30} $m>1.1$ &11.19&& 8.57&& 10.05&& 9.50\\ 
                                
                                & && && && \\
                                
                                Panel B &\multicolumn{7}{c}{2-SVJ Implied Volatility RMSE} \\
                                
                                & && && && \\
                                
                                \rowcolor{gray!30} $0.85 \leq m \leq1.1$ &11.97&& 11.19&& 13.21&& 8.67 \\
                                
                                 $m<0.85$ &30.39&& 19.81&& 10.57&& 19.98 \\
                                
                                \rowcolor{gray!30} $m>1.1$ &12.34&& 9.39&& 12.40&& 10.21\\          
                \hline               
            \end{tabular}                
        \end{small}                        
    \end{center}
    \caption{\label{tab:performance_moneyness_period} Pricing performances of the LHARG-ARJ and 2-SVJ model for different maturities and moneyness intervals.}
\end{table}
            
\section{Conclusions}\label{sec:CH2_conclusion}
Carbon markets are becoming increasingly vital for reducing emissions and boosting the transition to a low-emission economy. A quantitative detailed analysis of these markets is essential for market players and policymakers. 
In this paper, we analyzed Phase 3 of the EU ETS market using high-frequency data. First, we construct the realized volatility series disentangling continuous and discontinuous components. We then apply the HAR class model to estimate and forecast realized volatility, finding that incorporating jump and negative leverage improves the model's fit and forecasts. We then used these models to estimate multifactor stochastic volatility with jumps under historical measure $\mathbb{P}$ using the indirect inference method. Subsequently, we risk-neutralized the estimated models with a three-dimensional pricing kernel compensating for the equity and variance components' risks. We compared our results with the LHARG-ARJ model, obtaining comparable performances on the options sample. \\

This paper contributes to the quantitative understanding of the EU ETS during Phase 3. Our findings indicate that at least two factors are essential for capturing the carbon futures volatility dynamics. We also support the need to include a jump component in the return dynamics. Indeed, the calibration results suggest that including jumps slightly improves the pricing performance relative to a no-jump specification. Overall, our study provides valuable insights into the volatility and jump dynamics in carbon markets and the compensation for risk required by agents trading on the carbon options. Last, but not least, our approach offers a tractable framework for pricing carbon derivatives leveraging the information content of high-frequency trades.
\newpage

\appendix
\section{Appendices}
\subsection{Proof of results}\label{app:proofs}
\begin{proof}[Proof of \textbf{Proposition~\ref{Prop::pricing_Kernel}}]

  Given the price dynamics specified in Equation~(\ref{eq::phisical_SDE}) and the pricing kernel specified in Equation~(\ref{PricingKernel}), we have
    \begin{equation*}
    \log M_t - \log M_0 = \phi(\log(p_t) - \log(p_0)) + \int_0^t \delta\left(\sigma_{1,s}, \sigma_{2,s}\right) ds+\psi_1(\sigma_{1,t}^2-\sigma_0^2) + \psi_2(\sigma_{2,t}^2 - \sigma_{2,0}^2)\,.
\end{equation*}
Hence we have
\begin{equation*}
\begin{aligned}
d \log M_t = &~\phi d\log p_t + \delta\left(\sigma_{1,t}, \sigma_{2,t}\right)dt + \psi_1 d\sigma_{1,t}^2 + \psi_2 d\sigma_{2,t}^2 \\
 = & \left( \phi \mu\left(\sigma_{1,t} ,\sigma_{2,t}\right) + \delta(\sigma_{1,t}, \sigma_{2,t}) + \psi_1 m_1(\sigma_{1,t}) + \psi_2 m_2(\sigma_{2,t} )\right)dt + \phi \sigma_{1,t}dW_{1,t}^X +  \phi \sigma_{2,t}  dW_{2,t}^X +  \\
 &+  \psi_1 \Lambda_1 \sigma_{1,t} dW_{1,t}^{\sigma_1}  +  \psi_2 \Lambda_2 \sigma_{2,t}  dW_{2,t}^{\sigma_2}+ \phi c_X dJ_t\,.
\end{aligned}
\end{equation*}
Using Ito's Lemma we can derive the dynamics for the process $M_t$
\begin{equation*}
    d M_t = M_t d\log M^c_t + \frac{1}{2}M_t \Xi(\sigma_{1,t},\sigma_{2,t}) dt + M_t(e^{\phi c_X}-1)dJ_t\,,
\end{equation*}
where
\begin{equation*}
\Xi(\sigma_{1,t},\sigma_{2,t}) = \sigma_{1,t}^2\left(\phi^2 + 2\phi\rho_1\psi_1\Lambda_1 + \psi_1^2\right) + \sigma_{2,t}^2\left(\phi^2 + 2\phi\rho_2\psi_2\Lambda_2 + \psi_2^2\right)
\end{equation*}
and $M^c_t$ denotes the continuous part of $M_t$.
Now, define $Y_{1,t} = B_tM_t$, where $B_t = B_0 e^{rt}$ is the money market account process with $r$ the risk-free rate. The stochastic differential of $Y_{1,t}$ reads
\begin{equation*}
    \frac{dY_{1,t}}{Y_{1,t}} = \frac{dM_t}{M_t} + rdt\,.
\end{equation*}
To ensure the absence of arbitrage, $Y_{1,t}$ must be a martingale, meaning the drift is zero. We express this using the following notation $\frac{1}{dt}\mathbb{E}_t[Y_{1,t}] = 0$. We define the return risk premium as
\begin{equation*}
    \Tilde{\mu}(\sigma_{1,t}, \sigma_{2,t}) = \mu(\sigma_{1,t}, \sigma_{2,t}) - r + \frac{1}{2}(\sigma_{1,t}^2 + \sigma_{2,t}^2)\,.
\end{equation*}
Absence of arbitrage for $Y_{1,t}$ hence imply
\begin{equation*}
\delta(\sigma_{1,t}, \sigma_{2,t}) = -r(1 + \phi) - \phi \left( \tilde{\mu}(\sigma_{1,t},\sigma_{2,t}) - \frac{1}{2} (\sigma_{1,t}^2 + \sigma_{2,t}^2) \right) - \left(\psi_1 m_1(\sigma_{1,t}) + \psi_2 m_2(\sigma_{2,t})\right) - \frac{1}{2} \Xi - \lambda \mathbb{E}[e^{\phi c_X} - 1]\,. 
\end{equation*}
In the same way, consider $Y_{2,t} = p_t M_t$. By Ito's Lemma and using the result that $\mathbb{E}[dM_t/M_t]=-rdt$ we have
\begin{equation*}
    \frac{dY_{2,t}}{Y_{2,t}} = \frac{dM^c_t}{M_t} + \frac{dp^c_t}{p_t} + \left(\phi (\sigma_{1,t}^2 + \sigma_{2,t}^2) + \psi_1\Lambda_1 \rho_1 \sigma_{1,t}^2 +  \psi_2\Lambda_2 \rho_2 \sigma_{2,t}^2\right)dt + (e^{(\phi + 1)c_X} -1)dJ_t\,.
\end{equation*}
The absence of arbitrage implies that $\frac{1}{dt}\mathbb{E}_t[Y_{2,t}] = 0$ and we obtain the following equation
\begin{equation*}
    \Tilde{\mu}(\sigma_{1,t}, \sigma_{2,t}) = -\sigma_{1,t}^2(\phi + \psi_1\Lambda_1\rho_1) - \sigma_{2,t}^2(\phi + \psi_2\Lambda_2\rho_2)
    - \lambda\mathbb{E}[e^{\phi c_X}( e^{c_X}- 1)]\,.
\end{equation*}
Now consider $Y_{3,t} = M_t \Pi(t, p_t, \sigma_{1,t}^2, \sigma_{2,t}^2)$ where $\Pi(t, p_t, \sigma_{1,t}^2, \sigma_{2,t}^2)$ is the value at time $t$ of a traded asset with payoff $\Pi(p_T)$ at time $T$. Applying Ito's Lemma we get
\begin{equation*}
\begin{aligned}
    d\Pi_t = &\frac{\partial \Pi}{\partial t} dt + \frac{\partial \Pi}{\partial p} dp + \frac{1}{2} \frac{\partial^2 \Pi}{\partial p^2} (\sigma_{1,t}^2 + \sigma_{2,t}^2) p_t^2 dt+\frac{\partial \Pi}{\partial \sigma_{1}^2} d\sigma_{1,t}^2 +  \frac{\partial \Pi}{\partial \sigma_{2}^2} d\sigma_{2,t}^2 + \frac{\partial^2 \Pi}{\partial p \partial \sigma_{1}^2}p_t \rho_1 \Lambda_1 \sigma_{1,t}^2 dt \\
    &+  \frac{\partial^2 \Pi}{\partial p \partial \sigma_{2}^2} p_t \rho_2 \Lambda_2 \sigma_{2,t}^2 dt + \frac{1}{2} \frac{\partial^2 \Pi}{\partial \sigma_{1}^2} \Lambda_1^2 \sigma_{1,t}^2 dt + \frac{1}{2} \frac{\partial^2 \Pi}{\partial \sigma_{2}^2} \Lambda_2^2\sigma_{2,t}^2 dt \\
    &+ [\Pi(t,  p_t e^{c_X}, \sigma_{1,t}^2, \sigma_{2,t}^2) - \Pi(t, p_t, \sigma_{1,t}^2, \sigma_{2,t}^2)]dJ_t\,.
\end{aligned}
\end{equation*}
The dynamics of $Y_{3,t}$ can be specified in the following way
\begin{equation*}
\begin{aligned}
    \frac{dY_{3,t}}{Y_{3,t}} = &\frac{dM^c_t}{M_t} + \frac{d\Pi^c_t}{\Pi_t} + \frac{1}{\Pi_t } 
    \left(\frac{\partial \Pi}{\partial p} (\sigma_{1,t}^2(\phi + \psi_1\rho_1\Lambda_1)  + \sigma_{2,t}^2(\phi + \psi_2\rho_2\Lambda_2))p_t\right.\\
    & + \left.\frac{\partial \Pi}{\partial \sigma_{1}^2}\sigma_{1,t}^2 \Lambda_1(\phi \rho_1 + \psi_1\Lambda_1) + \frac{\partial \Pi}{\partial \sigma_{2}^2}\sigma_{2,t}^2 \Lambda_2(\phi \rho_2 + \psi_2\Lambda_2)\right) + c_{Y_{3,t}} dJ_t\,,
\end{aligned}
\end{equation*}
where \begin{equation*}
    c_{Y_{3,t}} = \frac{\Pi(t, p_t e^{c_X}, \sigma_{1,t}^2, \sigma_{2,t}^2)e^{\phi c_X} - \Pi(t,p_t, \sigma_{1}^2, \sigma_{2}^2)}{\Pi(t,p_t, \sigma_{1}^2, \sigma_{2}^2)}\,.
\end{equation*}
By setting $\frac{1}{dt}\mathbb{E}[dY_{3,t}/Y_{3,t}]=0$ and simplifying we get
\begin{equation*}
\begin{aligned}
    &-r\Pi(t, p_t, \sigma_{1,t}^2, \sigma_{2,t}^2) + \frac{\partial \Pi}{\partial t} + \frac{\partial \Pi}{\partial p}(r + \tilde{\mu}(\sigma_{1}^2, \sigma_2^2))p_t + \frac{\partial \Pi}{\partial \sigma_{1}^2}m_1(\sigma_{1,t}) + 
    \frac{\partial \Pi}{\partial \sigma_{2}^2}m_2(\sigma_{2,t}) + 
    \frac{1}{2} \frac{\partial^2 \Pi}{\partial p^2} (\sigma_{1,t}^2 + \sigma_{2,t}^2) p_t^2 \\
&+ \frac{1}{2}\frac{\partial^2 \Pi}{\partial \sigma_{1}^2}\Lambda^2_1\sigma_{1,t}^2 + \frac{1}{2}\frac{\partial^2 \Pi}{\partial \sigma_{2}^2}\Lambda^2_2 \sigma_{2,t}^2+ \frac{\partial^2 \Pi}{\partial p \partial \sigma_{1}^2} p_t \rho_1 \Lambda_1 \sigma_{1,t}^2  \\
    &+ \frac{\partial^2 \Pi}{\partial p \partial \sigma_{2}^2} p_t \rho_2 \Lambda_2 \sigma_{2,t}^2 + \frac{\partial \Pi}{\partial p} (\sigma_{1,t}^2 p_t(\phi + \psi_1\rho_1\Lambda_1)  + \sigma_{2,t}^2(\phi + \psi_2\rho_2\Lambda_2))\\
    & + \frac{\partial \Pi}{\partial \sigma_{1}^2}(\sigma_{1,t}^2 \Lambda_1(\phi \rho_1 + \psi_1\Lambda_1) + \frac{\partial \Pi}{\partial \sigma_{2}^2}(\sigma_{2,t}^2 \Lambda_2(\phi \rho_2 + \psi_2\Lambda_2))\\
    &+ \lambda \mathbb{E}[e^{\phi c_X}\left(\Pi(t,p_t e^{c_X}, \sigma_{1,t}^2, \sigma_{2,t}^2) - \Pi(t,p_t, \sigma_{1}^2, \sigma_{2}^2)\right)] = 0\,.
    \end{aligned}
\end{equation*}
Now, for every traded asset with price $\Pi(t, p_t, \sigma_{1,t}^2, \sigma_{2,t}^2)$ we have
\begin{equation*}
\begin{aligned}
    &-r\Pi( t, p_t, \sigma_{1,t}^2, \sigma_{2,t}^2) + \frac{\partial \Pi}{\partial t} + \frac{\partial \Pi}{\partial p}(r - \lambda^{*}\mathbb{E}[e^{c_X^{*}} -1 ])p_t + 
    \frac{\partial \Pi}{\partial \sigma_{1}^2} m^{*}_1 (\sigma_{1,t}) + \frac{\partial \Pi}{\partial \sigma_{2}^2} m^{*}_2(\sigma_{2,t}) \\
    &+ \frac{1}{2}\frac{\partial^2 \Pi}{\partial p^2} (\sigma_{1,t}^2 + \sigma_{2,t}^2)p_t^2 + \frac{1}{2}\frac{\partial^2 \Pi}{\partial \sigma_{1}^2} \Lambda^2_1 \sigma_{1,t}^2+ \frac{1}{2}\frac{\partial^2 \Pi}{\partial \sigma_{2}^2} {\Lambda^2_2} \sigma_{2,t}^2+ \frac{\partial^2 \Pi}{\partial p \partial \sigma_{1}^2}p_t \Lambda_1 \rho_1 \sigma_{1,t}^2 + \frac{\partial^2 \Pi}{\partial p \partial \sigma_{2}^2}p_t\Lambda_2 \rho_2 \sigma_{2,t}^2 \\
    &+ \lambda^* \mathbb{E}[\Pi(t,{\color{black}p_t e^{c_X^{*}}}, \sigma_{1,t}^2, \sigma_{2,t}^2) - \Pi(t,p_t, \sigma_{1}^2, \sigma_{2}^2)] = 0\,.
\end{aligned}
\end{equation*}
If we now compare the last two equations, we get
\begin{equation*}
    \begin{aligned}
        & m_1^{*}(\sigma_{1,t}) - m_1(\sigma_{1,t}) = \sigma_{1,t}^2(\phi \rho_1 \Lambda_1  + \psi_1 \Lambda_1^2) \\
        & m_2^{*}(\sigma_{2,t}) - m_2(\sigma_{2,t}) = \sigma_{2,t}^2( \phi \rho_2 \Lambda_2  + \psi_2 \Lambda_2^2) \\
        & {\color{black}\lambda^{*}\mathbb{E}[e^{c_X^*}-1] = \lambda \mathbb{E}[e^{\phi c_X}(e^{c_X}-1)]}\\
        & {\color{black}\lambda^{*}\mathbb{E}\left[\Pi\left(t, p_{t} e^{c_X^{*}}, \sigma_{1,t}^{2},\sigma_{2,t}^{2}\right)-\Pi\left(t, p_{t}, \sigma_{1,t}^{2},\sigma_{2,t}^{2}\right)\right]}\\
        & {\color{black}=\lambda \mathbb{E}\left[e^{\phi c_X}\left(\Pi\left(t, p_{t} e^{c_X}, \sigma_{1,t}^{2},\sigma_{2,t}^{2}\right)-\Pi\left(t, p_{t}, \sigma_{1,t}^{2},\sigma_{2,t}^{2}\right)\right)\right]}\,.
  \end{aligned}
\end{equation*}
From the last two relations, it follows that
$$
\lambda^{*} = \lambda \mathbb{E}[e^{\phi c_X}]
$$
and, for all $u \in \mathbb{R}$ 
\begin{equation*}
    \mathbb{E}[e^{iuc^*_X}] = \frac{\mathbb{E}\left[e^{iuc_X} (e^{\phi c_X})\right]}{\mathbb{E}\left[ e^{\phi c_X }\right]} =\exp{\left(\phi \sigma^2_X i u - \frac{1}{2}\sigma^2_X u^2 \right)}\,.
\end{equation*}
To conclude, the risk-neutral dynamics is given by
\begin{equation*}
\begin{aligned}
&d x_t=  {\left(r-q-\lambda^*\mathbb{E}[e^{c_X^*}-1]-\frac{1}{2}\left(\sigma_{1, t}^2+\sigma_{2, t}^2\right)\right) d t } +\sigma_{1, t}^2 d \Tilde{W}_{1,t}^{X}+\sigma_{2, t} d \Tilde{W}_{2, t}^{X}+c_X^* d \tilde{J}_t \\
&d \sigma_{1, t}^2= m_1^*(\sigma_{1,t})dt + \Lambda_1\sigma_{1,t} d\tilde{W}^{\sigma_1}_{1,t}\\
&d \sigma_{2, t}^2= m_2^*(\sigma_{2,t})dt + \Lambda_2\sigma_{2,t} d\tilde{W}^{\sigma_2}_{2,t}\,,
\end{aligned}
\end{equation*}
where we have
\begin{equation*}
    \begin{aligned}
        &\text{corr}(d \Tilde{W}_{1,t}^{X}, d\tilde{W}^{\sigma_1}_{1,t}) = \rho_1 \\
        &\text{corr}(d \Tilde{W}_{2,t}^{X}, d\tilde{W}^{\sigma_2}_{2,t}) = \rho_2\,.
    \end{aligned}
\end{equation*}
All other correlations, including those between the Poisson process and Brownian motions, are zero. \\
\end{proof}

\begin{proof}[Proof of \textbf{Lemma}~\ref{lem::chf}]
We derive the conditional CF of the 2-SVJ model, following the calculations in \cite{Duffie2000} and \cite{Heston++}. The CF takes the following form:
\begin{equation*}
\begin{aligned}
& \log \hat{f}_x(z ; \tau)=i\left(x_t+(r-q) \tau\right) z \\
& \quad+\sum_{j=1,2}\left(A_j^x(z ; \tau)+B_j^x(z ; \tau) \sigma_{j, t}^2\right)+C_{J}^x(z ; \tau),
\end{aligned}
\end{equation*}
where $\tau = T-t$ and the coefficients satisfy the following ODEs
\begin{equation*}
\begin{aligned}
& \frac{\partial A_j^x(z ; \tau)}{\partial \tau}=\kappa_j \omega_j B_j^x(z ; \tau) \\
& \frac{\partial B_j^x(z ; \tau)}{\partial \tau}=\frac{1}{2} \Lambda_j^2\left(B_j^x(z ; \tau)\right)^2-\left(\kappa_j-i z \rho_j \Lambda_j\right) B_j^x(z ; \tau)-\frac{1}{2} z(i+z) \\
& \frac{\partial C_{J}^x(z ; \tau)}{\partial \tau}=\lambda\left(\theta^{J}\left(z\right)-1-i \mathbb{E}[e^{c_X^*}-1] z\right)\,,
\end{aligned}
\end{equation*}
where $\theta^J (z)$ is the characteristic function of the jump sizes $c^*_X$. Given null initial condition at $\tau = 0$, the solution of this system of ODEs is
\begin{equation*}
\begin{aligned}
& A_j^x(z ; \tau)=\frac{\kappa_j \omega_j}{\Lambda_j^2}\left[\left(c_j-d_j\right) \tau-2 \log \left(\frac{1-g_j e^{-d_j \tau}}{1-g_j}\right)\right] \\
& B_j^x(z, \tau)=\frac{c_j-d_j}{\Lambda_j^2} \frac{1-e^{-d_j \tau}}{1-g_j e^{-d_j \tau}} \\
& C_{J}^x(z ; \tau)=\lambda \tau\left(\theta^{J}(z )-1-i \mathbb{E}[e^{c_X^*}-1] z\right)\,.
\end{aligned}
\end{equation*}
For simplicity, we have defined the following parameters:
\begin{equation*}
\begin{aligned}
c_j &= \kappa_j - i z \rho_j \Lambda_j \\
d_j &= \sqrt{c_j^2 + z (i + z) \Lambda_j^2}\\
g_j &= \frac{c_j - d_j}{c_j + d_j}\,.
\end{aligned}
\end{equation*}
The CF of nested models can be readily obtained from this result.
\end{proof}

\subsection{LHARG-ARJ estimated parameters and risk premia}\label{app:LHARGARJ}

This section reports the estimation results of the LHARG-ARJ model, obtained through likelihood maximization. For details, please refer to~\cite{Alitabetal}. Table~(\ref{tab:LHARG_ARJ_estimates}) presents the estimated parameter values and risk premia. 
                \begin{table}[h!]\setlength{\tabcolsep}{1.5em}
                    \begin{center}
                       
                            \begin{tabular}{lr} 
                                &    \multicolumn{1}{c}{}                           \\        
                                \hline
                                &               \\
                                Parameter & Estimates \\
                                \hline
                                &               \\
                                \rowcolor{gray!30} $\Phi_c$           &   3.0 (9e-01)                  \\	
                                $\Phi_j$           &   -61 (5)                   \\	
                                \rowcolor{gray!30} $\theta$                     &   9.3e-05 (3e-06)             \\ 
                                $\kappa$     	     & 2.24                     \\ 
                                \rowcolor{gray!30} $\beta_{d} $	 & 4.2e+03 (3e+02)              \\ 
                                $\beta_{w}$      & 2.9e+03 (4e+02)              \\ 
                                \rowcolor{gray!30} $\beta_{m}$      & 6e+02 (3e+02)              \\ 
                                $\alpha_{d} $ 	         & 0.62 (0.01)              \\ 
                                \rowcolor{gray!30} $\alpha_{w} $	   & 0.42 (0.15)              \\ 
                                $\alpha_{m} $	      & 0.17 (0.37)              \\ 
                                \rowcolor{gray!30} $\gamma $          & 5 (3)              \\ 
                                $\bar{\lambda}$        &  2.2e-02                \\
                                \rowcolor{gray!30} $\xi$  & 0.84 (0.02)             \\
                                $\zeta$ & 0.12 (0.02)              \\
                                \rowcolor{gray!30} $\mu_J$ & -7.9e-03 (7e-04)         \\
                                $\sigma_J$ & 8.5e-03 (6e-04)             \\
                                &  \\
                                 & Risk premia \\
                                \rowcolor{gray!30} $\nu_c$ &-6.72e+02 \\
                                $\nu_j$ & -1.99e+03 \\                      
                                & \\
                                & Log-likelihood \\
                                \rowcolor{gray!30} $L^y$            &  2890              \\
                                $L^\mathrm{CRV}$ & -10258              \\
                                \rowcolor{gray!30} $L^\mathrm{JRV}$ &   1951              \\
                                Persistence $\mathrm{CRV}_t$       & 0.759               \\
                                \rowcolor{gray!30} Persistence $\omega_t$      & 0.968               \\
                                \hline
                            \end{tabular}
                    
                    \end{center}
                    \hspace{-1.5cm}
               
                        \caption{ Maximum likelihood estimates (standard errors in parenthesis) for the LHARG-ARJ model on ECF Phase 3. Risk premia are calibrated from a sample of futures put and call options. }\label{tab:LHARG_ARJ_estimates}
                 
                \end{table}
Parameters are in daily not percentage units. The results are qualitatively similar to those found fitting the LHAR-CJ model. The coefficients $\beta_{\{d,w,m\}}$ associated with the continuous component are the most significant, particularly the daily and weekly coefficients. The leverage component, described by the parameters $\alpha_{\{d,w,m\}}$, shows similar results for daily and weekly but the monthly coefficient exhibits the largest error. Overall, the leverage effect does not appear to have a large magnitude.
The drift parameters $\Phi_c$ and $\Phi_j$, corresponding to the continuous and discontinuous components, are significant. Notably, the latter is large and negative. Regarding the jump size distribution, the mean $\mu_J$ is negative, and the standard deviation $\sigma_J$ is of the same order of magnitude. This suggests that jumps are primarily associated with negative shocks. In comparison to the original study on S\&P 500 data of \cite{Alitabetal}, we observe a lower persistence of volatility, while the jump persistence shows a slightly lower value. The jump intensity is higher, while the leverage component is weaker. These parameters are estimated under the historical measure.\\ 
The pricing kernel for the LHARG-ARJ depends parametrically on four risk premia, two of which are determined by the arbitrage conditions, as outlined in \cite{Alitabetal}. In the table, we present the calibrated variance risk premia associated with the continuous, $\nu_c$, and discontinuous, $\nu_j$, components of the quadratic variation.
As expected, both risk premia are large and negative.

\newpage

\begin{center}
    ONLINE SUPPLEMENTAL INFORMATION
\end{center}

\subsection*{The LHAR-CJ model}
\begin{figure}[h!]
    \centering
    \includegraphics[width =0.9 \textwidth]{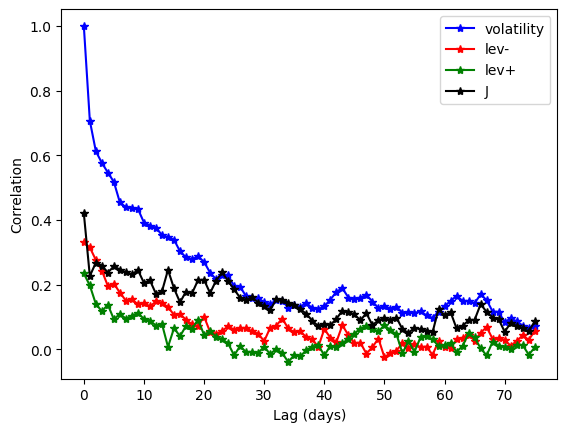}
    \caption{Lagged correlation function between past values ($RV_{t-h}$) and current daily realized volatility estimates ($RV_t$) as a function of $h$, along with negative returns $r_{t-h}^-$, positive returns $r_{t-h}^+$, and jumps quadratic variation ($\hat{J}_{t-h}$).}
    \label{fig::Lagged Corre}
\end{figure}

For the ease of the reader, this section of the online SI mainly recalls results from \cite{CorsiRenoJBES}.
We work in a filtered probability space $(\Omega, (\mathcal{F}_t)_{t \in [0,T]}, \mathcal{F}, \mathbb{P})$. We assume that the logarithmic price of the European Carbon Futures (ECF) is noted as  $X_t$ and satisfy the following assumption
\begin{assumption}\label{AssumptionProcess}
    $(X_t)_{t \in [0,T]}$ is a real-values process and can be put in the form of an Ito semimartingale:
    \begin{equation*}
    dX_t = \mu_tdt + \sigma_t dW_t +dN_t,
\end{equation*}
where $\mu_t$ is predictable, $\sigma_t$ is càdlag and $W_t$ is the standard Brownian motion. The jump part is $dN_t = c_{X_t}dJ_t$, where $J_t$ is a non-explosive Poisson process whose intensity is an adapted stochastic process $\lambda_t$ and $c_{X_t}$ is an adapted random variable that measures the size of the jump at time $t$. Size and time of a jump are i.i.d random variables and we have $\mathbb{P}(\{c_{X_t}=0\})=0$, $\forall t \in [0,T].$
\end{assumption}
\noindent We set our time window $T=1$ day and we denote the daily close-to-close return as $r_t$. We study the quadratic variation of the process within this period. The quadratic variation of the process $X_t$ is defined in the following way
\begin{equation*}
    [X]_t^{t+T} = \int _t ^ {t+T} \sigma^2_s ds+ \sum _{j = N_t} ^ {N_{t+T}} c_{X_j}^2 
\end{equation*}
We define the continuous component as $[X^c]_t^{t+T} = \int _t ^ {t+T} \sigma^2_s \, ds$ and the discontinuous component as $[X^d]_t^{t+T} = \sum _{j = N_t} ^ {N_{t+T}} c_j^2$. These quantities are not directly observable and we need to use consistent estimators. We divide the daily time interval $[t, t+T]$ into $n$ evenly spaced sub-intervals of length $\delta = T/n$. On this grid, we have evenly sampled returns defined in this way
\begin{equation*}
    \Delta_{j,t}X = X_{j\delta + t } - X_{(j-1)\delta + t }, \quad j=1,\ldots , n
\end{equation*}
From now on, for simplicity of notation, we only write $\Delta_{j}X$ to refer to this quantity. The quantity $\delta$ basically defines the frequency for calculating the intra-day returns. The most important estimator of $[X]_t^{t+T}$ is the realized variance (see \cite{Shepard::RV}), defined as:
\begin{equation*}
     RV_{\delta}(X)_t = \sum _{j=1} ^n (\Delta _j X)^2,
\end{equation*}
which converges in probability to  $[X]_t^{t+T}$ as $\delta \longrightarrow 0$.
We define the notations for the realized estimators of the quadratic variation components as follows: $\hat{V}_t$ represents the estimator for $[X]_{t}^{t+T}$, $\hat{C}_t$ for $[X^c]_{t}^{t+T}$, and $\hat{J}_t$ for $[X^d]_{t}^{t+T}$. 
To estimate the total quadratic variation, we use 5-minute intraday returns. While other methods, such as \cite{zhang2s} two-scale estimator, can estimate quadratic variation using tick-by-tick data, we find that 5-minute intervals are appropriate given the market's limited liquidity at times. To differentiate between continuous and discontinuous components, we employ the \textit{C-Tz} test developed by \cite{CORSIJump}, preceded by data pre-treatment: we first calculate Jump-Adjusted returns as described by \cite{Andersen2010}, followed by a threshold-based trimming technique outlined in \cite{CorsiFusVE}. This procedure is essential for constructing the time series of both continuous and discontinuous components of quadratic variation, as well as estimating the number of intraday jumps $n_t$ and their sizes $c_{X_{t}}$. These quantities enable us to define the LHARC-CJ model class and construct the likelihood for estimating the LHARG-ARJ pricing model used as a benchmark model.
The estimators of $[X^c]_{t}^{t+T}$ and $[X^d]_{t}^{t+T}$, along with further details on this procedure, are available in the next section.
The LHAR-CJ model is constructed by combining heterogeneity in realized volatility, leverage, and jumps, utilizing daily, weekly, and monthly frequencies. Figure~(\ref{fig::Lagged Corre}) illustrates components of the LHAR-CJ model by showing the lagged correlation function between Realized Volatility $RV_t$ and $RV_{t-h}$ for the ECF futures series of Phase 3, alongside its correlation to negative returns, positive returns, and the jump component. The autocorrelation of realized volatility decays slowly, a well-known characteristic. The lagged correlation between $RV_t$ and negative returns shows the presence of a leverage effect, though it is relatively weak. The correlation with positive returns is even weaker, leading us to focus on negative returns for our analysis. In contrast, the jump component positively influences $RV_t$ and decays more slowly than the returns component.
We use the variables specified in logarithmic scale as common practice, defining the averaged aggregated variables over $h$ days as follows (with jumps aggregated only):
\begin{equation*}
\begin{aligned}
\log \widehat{\mathrm{V}}_t^{(h)} & =\frac{1}{h} \sum_{j=1}^h \log \widehat{\mathrm{V}}_{t-j+1}, \quad \log \widehat{\mathrm{C}}_t^{(h)}=\frac{1}{h} \sum_{j=1}^h \log \widehat{\mathrm{C}}_{t-j+1}, \\
r_t^{(h)} & =\frac{1}{h} \sum_{j=1}^h r_{t-j+1}, \quad \widehat{\mathrm{J}}_t^{(h)}=\sum_{j=1}^h \widehat{\mathrm{J}}_{t-j+1} .
\end{aligned}
\end{equation*}
To model the negative leverage effect, we define \( r_t^{(h)-} = \min(r_t^{(h)}, 0) \). 
We define the LHAR-CJ with the standard negative leverage effect as follows:
\begin{equation*}
\begin{aligned}
\log \widehat{\mathrm{V}}_{t+h}^{(h)}=c & +\beta^{(d)} \log \widehat{\mathrm{C}}_t+\beta^{(w)} \log \widehat{\mathrm{C}}_t^{(5)}+\beta^{(m)} \log \widehat{\mathrm{C}}_t^{(22)} \\
& +\alpha^{(d)} \log \left(1+\widehat{\mathrm{J}}_t\right)+\alpha^{(w)} \log \left(1+\widehat{\mathrm{J}}_t^{(5)}\right)+\alpha^{(m)} \log \left(1+\widehat{\mathrm{J}}_t^{(22)}\right) \\
& +\gamma^{(d)} r_t^{-}+\gamma^{(w)} r_t^{(5)-}+\gamma^{(m)} r_t^{(22)-}+\varepsilon_t^{(h)}
\end{aligned}
\end{equation*}
The parameters $\{ c, \beta^{(d, w, m)}, \alpha^{(d, w, m)}, \gamma^{(d, w, m)} \}$ along with i.i.d noise $\varepsilon_t^{(h)}$ characterize the model. This formulation incorporates well-known models for modeling and forecasting realized volatility. When $\alpha^{(d, w, m)} = \gamma^{(d, w, m)} = 0$, the leverage and jump components are omitted, leading to $\hat{C}_t = \hat{V}_t$, thus simplifying to the well-known HAR model of \cite{CorsiRV}. If there is no separation of quadratic variation (i.e., $\alpha^{(d, w, m)}=0$), the model is identified as the LHAR model. When $\gamma^{(d, w, m)}=0$, it corresponds to the HAR-CJ model proposed by \cite{Andersen2007}, which treats continuous and discontinuous components as separate explanatory variables.
We estimate the LHAR-CJ and its variants using ordinary least squares (OLS) with Newey-West covariance correction. 

\subsection*{Realized measures and data pre-treatment}

This section outlines the estimators for both continuous and discontinuous components of realized volatility and the jump detection test. This section mainly refers to the work of \cite{CORSIJump}.

\noindent \cite{BardoffNielsenBPV} introduced the multipower variation estimator to separate the continuous and discontinuous parts of quadratic variation, defined as follows
\begin{equation*}
    \text{MPV}_{\delta}(X)_t ^ {[\gamma_1,\ldots,\gamma_M]} = \delta^{1-\frac{1}{2}(\gamma_1 + \ldots \gamma_M)} \sum_{j=M}^{[T/\delta]} | \Delta_{j-k+1}X|^{\gamma_k}.
\end{equation*}
As $\delta \to 0$, this quantity converges in probability to $\mu_{\gamma}\int _{t} ^ {t+T} \sigma_s^{\gamma_1 +\ldots + \gamma_M} ds$ for a suitable constant $\mu_{\gamma}$. Asymptotic properties of this estimator were analyzed in \cite{AsymptoticBPV} and \cite{AsymptoticBPVJUMP}. In practical applications, multipower variation is typically employed to estimate the continuous component of quadratic variation, resulting in the well-known bipower variation, defined as:
\begin{equation*}
    \text{BPV}_{\delta}(X)_t = \mu_1^{-2}\text{MPV}_{\delta}(X)_t^{[1,1]} = \mu_1^{-2} \sum _{j=2} ^ {[T/\delta]} |\Delta_{j-1}X|\cdot|\Delta_{j}X|,
\end{equation*}
where the constant $\mu_1 \simeq 0.7979$ and we have convergence in probability to $[X^c]_{t}^{t+T}$ as $\delta \longrightarrow 0$.
To detect a jump on a specific day, we need to check for the presence of a discontinuity. We use the method from \cite{CORSIJump} for the \textit{C-Tz} test, a modified version of the \textit{z-test} by \cite{BNS_Ztest}. The original test assesses the difference between the RV and BPV but suffers from finite-sample bias. To mitigate this, a strictly positive threshold function, $\tau_s : [t, t+T] \longrightarrow \mathbb{R}^+$, is introduced, satisfying the conditions outlined by \cite{Mancini}.
\begin{equation*}
    \lim _{\delta \xrightarrow{} 0} \tau(\delta) = 0, \quad \lim _{\delta \xrightarrow{} 0} \frac{\delta \log \frac{1}{\delta}}{\tau(\delta)} = 0.
\end{equation*}
In practice, the threshold function must decrease slower than the modulus of continuity of Brownian motion to ensure convergence in probability. This can be accomplished using a multiple of an estimator for the local spot variance:
\begin{equation*}
    \tau_t = c_\tau^2 \cdot \hat{\mathcal{V}_t},
\end{equation*}
where $c_\tau$ is a scaling constant and $\hat{\mathcal{V}_t}$ can be estimated recursively as in \cite{CorsiRenoJBES}. To define the \textit{C-Tz} test we need to use the set of estimators called threshold multipower variation introduced by \cite{CORSIJump}, defined as:
\begin{equation*}
\text{TMPV}_\delta(X)_t^{\left[\gamma_1, \ldots, \gamma_M\right]}=\delta^{1-\frac{1}{2}\left(\gamma_1+\ldots+\gamma_M\right)} \sum_{j=M}^{[T / \delta]} \prod_{k=1}^M\left|\Delta_{j-k+1} X\right|^{\gamma_k} I_{\left\{\left|\Delta_{j-k+1} X\right|^2 \leq \vartheta_{j-k+1}\right\}}.
\end{equation*}
The indicator function $I_{(\cdot)}$ corrects the bias in multipower variation caused by consecutive jumps. Additionally, to develop a test for detecting jumps with these estimators, we must adjust for the finite-sample negative bias introduced by $\delta$. This is achieved by defining a class of corrected threshold estimators used to construct the \textit{C-Tz }statistic. We define the following quantity
\begin{equation*}
\text{C-TMPV}_\delta^{\left[\gamma_1, \ldots, \gamma_M\right]}=\delta^{1-\frac{1}{2}\left(\gamma_1+\ldots+\gamma_M\right)} \sum_{j=M}^{[T / \delta]} \prod_{k=1}^M Z_{\gamma_k}\left(\Delta_{j-k+1} X, \vartheta_{j-k+1}\right)
\end{equation*}
where the function $Z_{\gamma}(x, y)$ is defined as:
\begin{equation*}
Z_\gamma(x, y)= \begin{cases}|x|^\gamma & \text { if } x^2 \leq y \\ \frac{1}{2 N\left(-c_{\vartheta}\right) \sqrt{\pi}}\left(\frac{2}{c_{\vartheta}^2} y\right)^{\frac{\gamma}{2}} \Gamma\left(\frac{\gamma+1}{2}, \frac{c_{\vartheta}^2}{2}\right) & \text { if } x^2>y\end{cases}
\end{equation*}
The C-Tz is then defined in the following way
\begin{equation*}
\text{C-Tz}=\delta^{-\frac{1}{2}} \frac{\left(\text{RV}_\delta-\text{C-TBPV}_\delta\right) \cdot \text{RV}_\delta^{-1}}{\sqrt{\left(\frac{\pi^2}{4}+\pi-5\right) \max \left\{1, \frac{\text{C-TTriPV}_\delta}{\left(\text{C-TBPV}_\delta\right)^2}\right\}}},
\end{equation*}
where $\text{C-TBPV}_\delta = \text{C-TMPV}_\delta^{\left[1,1\right]}$, and  $\text{C-TTriPV}_\delta = \text{C-TMPV}_\delta^{\left[4/3, 4/3, 4/3\right]}$, for more details, see \cite{CORSIJump}.
Under these assumptions, if \( dJ_t = 0 \), then \textit{C-Tz} converges to \( \mathcal{N}(0,1) \) in law as \( \delta \to 0 \). For a significance level \( \alpha \), we assess the daily jump component's statistical significance by comparing the \textit{C-Tz} statistic to the standard normal quantile \( \Phi_{1-\alpha} \). If \( \textit{C-Tz} > \Phi_{1-\alpha} \), we reject the null hypothesis. Using the Threshold Bipower Variation (\( \text{TBPV}_t \)) for estimating the continuous component of quadratic variation\footnote{Defined as $\text{TBPV}_t = \text{TMPV}_\delta(X)_t^{[1,1]}$}, when a jump is detected, we attribute the difference \( \mathrm{RV}_t - \mathrm{TBPV}_t \) to the jump component: \( \hat{J}_t = \mathbb{I}_{\mathrm{C-Tz}_t > \Phi_{1-\alpha}}(\mathrm{RV}_t - \mathrm{TBPV}_t) \). Specifically, we set \( \hat{V}_t = \hat{C}_t = \mathrm{RV}_t \) and \( \hat{J}_t = 0 \) on days when we do not reject the null hypothesis. If the test rejects the null, we set \( \hat{C}_t = \text{TBPV}_t \) and \( \hat{J}_t = \max(\text{RV}_t - \text{TBPV}_t , 0)\). The $\hat{C}_t$ series is then cleaned by removing extreme observations likely caused by volatility jumps, using a threshold-based jump detection method as outlined by \cite{CorsiFusVE}. Finally, since our volatility estimator is based on returns during the trading period (from market open to close), we rescale it to align with the unconditional mean of squared daily returns (close-to-close), accounting for overnight returns as well.
The \textit{C-Tz} identifies days with at least one jump but does not indicate the number of intraday jumps. To determine the actual number of jumps, we follow the iterative procedure of \cite{Andersen2010}, using as outlined before $\delta = \text{5 minutes}$. Upon identifying a day with a jump, we remove the largest 5-minute return and replace it with the day's average return. We then repeat the \textit{C-Tz} test on the adjusted series. If the test fails to reject the null hypothesis, we conclude that only one jump occurred. If the null is rejected, we identify another jump and repeat the process until the null hypothesis is no longer rejected, resulting in a series of intraday 5-minute jump returns. This method allows us to non-parametrically recover the time series of daily jump counts, identified by \(n_t\), and jump sizes, \(c_{X_{t,i}}\). If the quantity \(\sum_{i=1}^{n_t} |c_{X_{t,i}}|^2\) does not match \(\mathbb{I}_{\mathrm{C-Tz}_t>\Phi_{1-\alpha}}\left(\mathrm{RV}_t-\mathrm{TBPV}_t\right)\), we scale \(c_{X_{t,i}}\) to ensure consistency. With the number and size of intraday jumps established, their contribution to the total daily return is easily calculated. The daily jump-adjusted return series is obtained by subtracting \(\sum_{i=1}^{n_t} c_{X_{t,i}}\) from the daily returns. A key advantage of the iterative non-parametric method by \cite{Andersen2010} is that both the number of intraday jumps, \(n_t\) and the size of the jumps, \(c_{X_{i,t}}\), become observable quantities and can be directly used in the estimation of the LHARG-ARJ model parameters via maximum likelihood, for more details refer to \cite{Alitabetal}. 

\subsection*{The indirect inference estimation technique}

While volatility analysis is typically conducted in discrete time, the option pricing exercise we discuss is framed in continuous time.
This section provides an overview of the indirect inference method developed by \cite{Gourioeux1993} to link continuous-time models with HAR models (see \cite{CorsiRenoJBES} for an application to the S\&P of this methodology).
Indirect inference is a simulation-based method for estimating the parameters of a structural model, consisting of two stages. First, an auxiliary model is fitted to the observed data. Next, a binding function -- either analytical or simulated -- maps the structural model parameters to the auxiliary statistic. Indirect inference then calibrates the structural model parameters to minimize the distance between the estimated parameters of the auxiliary models.
The structural model we use is from the class of affine multifactor models with jumps that will be presented in the next section. The auxiliary model is the LHAR-CJ and the nested models HAR and LHAR. Denote by $\hat{\beta}_T$ the parameter vector of the auxiliary model estimated on the data and $\theta$ the parameter vector of the structural model. Then, for a given $\theta$, we simulate $S$ simulated replicas of the structural model with a fixed intraday frequency $\delta$. We estimate the series of the realized quantities on this simulated series and then we estimate the auxiliary model on each replica. We denote these estimates by $\hat{{\beta}}^s_T(\theta)$, with $s = 1, \ldots, S$. The structural parameter vector is then estimated by minimizing the following quantity 
\begin{equation*}
\hat{\theta}_{ST} = \argmin_{\theta} \chi^2_T(\theta)
\end{equation*}
where
\begin{equation*}
 \chi^{2} = \left(\hat{\beta}_T - \frac{1}{S} \sum_{s=1} ^ S \hat{\beta}_T^s (\theta)\right)^{'}\Omega_T\left(\hat{\beta}_T - \frac{1}{S} \sum_{s=1} ^ S \hat{\beta}_T^s (\theta)\right),
\end{equation*}
where $\Omega_T$ denote a suitable positive-definite weighting matrix, that can be set as the variance-covariance matrix of the auxiliary model parameters estimated from the data. The implied parameters of the auxiliary models are the average of $\hat{\beta}_T^s(\hat{\theta}_{ST})$. In this way, the continuous-time model estimated captures the stylized facts described by the discrete-time model.
Further details on the indirect inference estimator and its asymptotic properties can be found in the \cite{Gour1996}.

\subsection*{Additional Analyses}

\begin{figure}[h!]
    \centering
    \includegraphics[width = \textwidth]{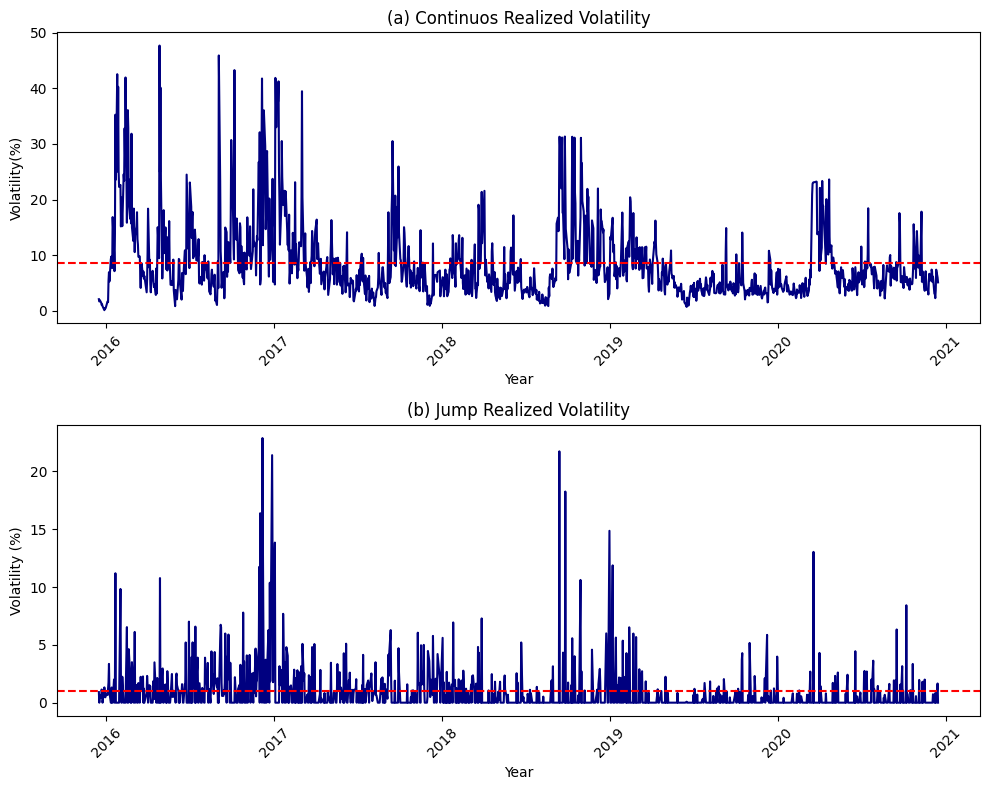}
    \caption{Realized quantities with mean value marked with dashed red line: (a) Continuous realized volatility time series ($\hat{C}_t)$. (b) Jump RV time series ($\hat{J}_t$).\label{fig::ContJUmp}}
\end{figure}

\begin{figure}[h!]
    \centering
    \includegraphics[width = \textwidth]{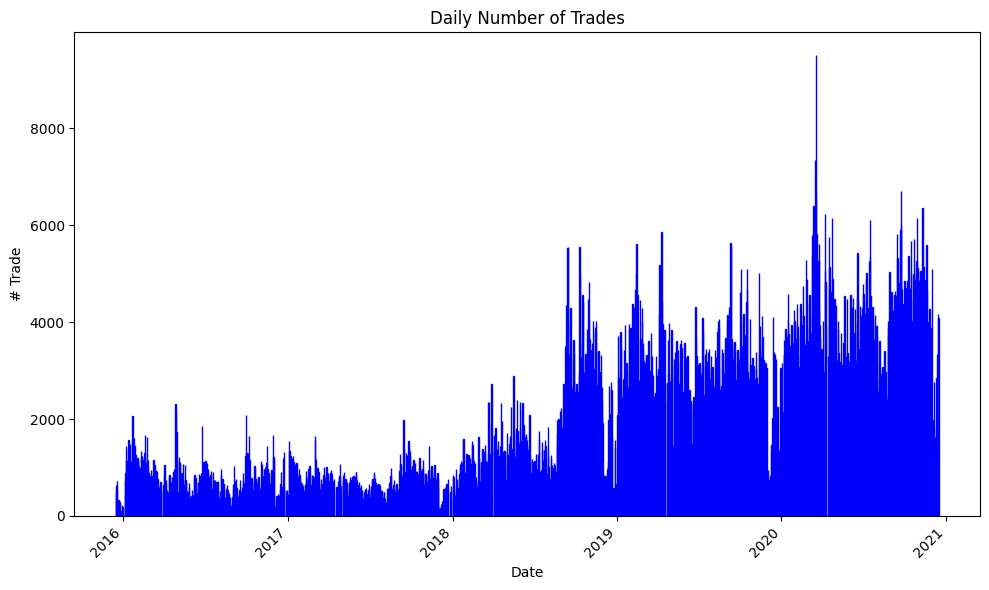}
    \caption{Daily EU-ETS trading volume during the analyzed period.}\label{fig::TradeNum}
\end{figure}

This section provides further analysis of the dataset and the key quantities involved. Figure~(\ref{fig::ContJUmp}) illustrates the estimated continuous and discontinuous components of quadratic variation obtained after data pre-treatment, which are used as covariates in the LHAR-CJ model together with leverage effect. The jump variation has stronger jump intensity and size at the beginning of the dataset, with some extreme values clustered during periods of high volatility. The continuous component mirrors the total variation shown in Figure~(\ref{fig::RVdwm}). The mean of the continuous component, $\hat{C}_t$, is 8.6\%, while the mean of $\hat{J}$ is 0.97\%.
We now analyze intraday liquidity during Phase 3 of the EU-ETS (from December 2015). Figure~(\ref{fig::TradeNum}) shows a consistent increase in market liquidity after 2018. Prior to this, daily trades remained mostly below 2000. Overall, the dataset averages 2100 daily trades, with a peak of 9500 during the 2020.
This low liquidity during the early years of the dataset could also be the cause of high realized volatility during market events.\\
We aim to further explore the leverage effect and its implications. As noted in Section~\ref{sec::Empirical App}, while the leverage effect in the LHAR model improves the fit, the associated coefficients exhibit low $t$-statistics, with the monthly component being particularly insignificant. We opted for negative leverage as it demonstrates superior in-sample statistics compared to positive leverage, as shown in Table~(\ref{tab:Lev1}).\\
\begin{table}[h!]
\centering
\begin{tabular}{l|ccc}
\toprule
Model & AIC&BIC& $R^{2}_{\textit{adj}}$ \\
\midrule
\textbf{LHAR-CJ-} & 1407&1459  &           0.622   \\
\text{LHAR-CJ+} &  1431&1482&                 0.610  \\
\textbf{LHAR-} &     1407&            1443&   0.618 \\
\text{LHAR+}  &      1419&1455&               0.607 \\
\bottomrule
\end{tabular}\caption{Leverage analysis for LHAR-CJ models. \label{tab:Lev1}}
\end{table}
\noindent We can further analyze the leverage component in this market by fitting a GJR-GARCH model to the close-to-close returns $r_t$ derived from tick-by-tick data. The model is specified as following
\begin{equation*}
    \begin{aligned}
        r_t &= \epsilon_t \\
        \sigma^2_t & = \omega + \alpha_{t-1} \epsilon^2_{t-1} + \gamma \epsilon^2_{t-1}\mathbbm{1}_{\{\epsilon_{t-1} < 0\}} + \beta \sigma^{2}_{t-1}\\
        \epsilon_t & = \sigma_t e_t, \quad , e_t \sim \mathcal{N}(0,1)
    \end{aligned}
\end{equation*}
Our estimates, shown in Table~(\ref{tab:garch_results}), indicate that the persistence parameter $\beta$ is the most significant, while the leverage parameter $\gamma$ is positive, suggesting a negative leverage effect but with a low $t$-statistics. Thus, we conclude that the negative leverage component exists in this market, but with a small magnitude.
\begin{table}[h!]
\centering
\begin{tabular}{lcccc}
\toprule
Parameter & Coefficient & Std. Error & $t$-statistics & $p$-Value \\
\midrule
\hspace{0.3cm} $\omega$ & 0.877  & 0.300 & 2.92 & 0.003 \\
\hspace{0.3cm} $\alpha$ & 0.105 & 0.035 & 2.94 & 0.003 \\
\hspace{0.3cm} $\gamma$ & 0.090 & 0.065 & 1.53 & 0.097 \\
\hspace{0.3cm} $\beta$ & 0.767 & 0.054 & 14.01 & 0.000 \\
\bottomrule
\end{tabular}\caption{GJR-GARCH model estimation results. \label{tab:garch_results}}
\end{table}

\begin{thebibliography}{}

\bibitem[Alberola et~al., 2007]{Alberola2}
Alberola, E., Chevallier, J., and Chèze, B. (2007).
\newblock European carbon prices fundamentals in 2005-2007: The effects of energy markets, temperatures and sectorial production.
\newblock EconomiX Working Papers 2007-33, University of Paris Nanterre, EconomiX.

\bibitem[Alberola et~al., 2008]{Alberola3}
Alberola, E., Chevallier, J., and Chèze, B. (2008).
\newblock Price drivers and structural breaks in european carbon prices 2005–2007.
\newblock {\em Energy Policy}, 36:787--797.

\bibitem[Alitab et~al., 2020]{Alitabetal}
Alitab, D., Bormetti, G., Corsi, F., and Majewski, A.~A. (2020).
\newblock {A jump and smile ride: Jump and variance risk premia in option pricing}.
\newblock {\em Journal of Financial Econometrics}, 18(1):121--157.

\bibitem[Andersen et~al., 2007]{Andersen2007}
Andersen, T., Bollerslev, T., and Diebold, F. (2007).
\newblock Roughing it up: Including jump components in the measurement, modeling, and forecasting of return volatility.
\newblock {\em The Review of Economics and Statistics}, 89(4):701--720.

\bibitem[Andersen et~al., 2010]{Andersen2010}
Andersen, T., Bollerslev, T., Frederiksen, P., and Nielsen, M. (2010).
\newblock Continuous-time models, realized volatilities, and testable distributional implications for daily stock returns.
\newblock {\em Journal of Applied Econometrics}, 25(2):233--261.

\bibitem[Bakshi et~al., 1997]{Bakshi}
Bakshi, G., Cao, C., and Chen, Z. (1997).
\newblock Empirical performance of alternative option pricing models.
\newblock {\em The Journal of Finance}, 52(5):2003--2049.

\bibitem[Bandi and Renò, 2016]{BANDIReno}
Bandi, F. and Renò, R. (2016).
\newblock Price and volatility co-jumps.
\newblock {\em Journal of Financial Economics}, 119(1):107--146.

\bibitem[Barndorff-Nielsen et~al., 2006a]{AsymptoticBPV}
Barndorff-Nielsen, O.~E., Graversen, S.~E., Jacod, J., and Shephard, N. (2006a).
\newblock Limit theorems for bipower variation in financial econometrics.
\newblock {\em Econometric Theory}, 22(4):677--719.

\bibitem[Barndorff-Nielsen and Shephard, 2002]{Shepard::RV}
Barndorff-Nielsen, O.~E. and Shephard, N. (2002).
\newblock Econometric analysis of realized volatility and its use in estimating stochastic volatility models.
\newblock {\em Journal of the Royal Statistical Society. Series B (Statistical Methodology)}, 64(2):253--280.

\bibitem[Barndorff-Nielsen and Shephard, 2004]{BardoffNielsenBPV}
Barndorff-Nielsen, O.~E. and Shephard, N. (2004).
\newblock {Power and Bipower Variation with Stochastic Volatility and Jumps}.
\newblock {\em Journal of Financial Econometrics}, 2(1):1--37.

\bibitem[Barndorff-Nielsen and Shephard, 2005]{BNS_Ztest}
Barndorff-Nielsen, O.~E. and Shephard, N. (2005).
\newblock {Econometrics of Testing for Jumps in Financial Economics Using Bipower Variation}.
\newblock {\em Journal of Financial Econometrics}, 4(1):1--30.

\bibitem[Barndorff-Nielsen et~al., 2006b]{AsymptoticBPVJUMP}
Barndorff-Nielsen, O.~E., Shephard, N., and Winkel, M. (2006b).
\newblock {Limit theorems for multipower variation in the presence of jumps}.
\newblock {\em Stochastic Processes and their Applications}, 116(5):796--806.

\bibitem[Baschetti et~al., 2022]{SINCBasch}
Baschetti, F., Bormetti, G., Romagnoli, S., and Rossi, P. (2022).
\newblock The {SINC} way: A fast and accurate approach to {F}ourier pricing.
\newblock {\em Quantitative Finance}, 22:427--446.

\bibitem[Bates, 2000]{Bates2000}
Bates, D.~S. (2000).
\newblock Post-'87 crash fears in the {S}\&{P} 500 futures option market.
\newblock {\em Journal of Econometrics}, 94(1-2):181--238.

\bibitem[Bates, 2015]{Bates}
Bates, D.~S. (2015).
\newblock {Jumps and Stochastic Volatility: Exchange Rate Processes Implicit in Deutsche Mark Options}.
\newblock {\em The Review of Financial Studies}, 9(1):69--107.

\bibitem[Benschop and L{\'o}pez-Cabrera, 2017]{BenschCabrera}
Benschop, T. and L{\'o}pez-Cabrera, B. (2017).
\newblock Realized volatility of {CO}$_2$ futures.

\bibitem[Benth et~al., 2017]{BenthCarbon}
Benth, F.~E., Eriksson, M., and Westgaard, S. (2017).
\newblock {Stochastic Volatility Modeling of Emission Allowances Futures Prices in the European Union Emission Trading System Market}.
\newblock In Secomandi, N., editor, {\em {Real Options in Energy and Commodity Markets}}, World Scientific Book Chapters, chapter~3, pages 63--115. World Scientific Publishing Co. Pte. Ltd.

\bibitem[Black, 1976]{black76}
Black, F. (1976).
\newblock The pricing of commodity contracts.
\newblock {\em Journal of Financial Economics}, 3(1-2):167--179.

\bibitem[Borghesi and Flori, 2019]{EUETSBrexit}
Borghesi, S. and Flori, A. (2019).
\newblock With or without {U}({K}): A pre-{B}rexit network analysis of the {EU} {ETS}.
\newblock {\em PLOS ONE}, 14(9):1--17.

\bibitem[Carmona and Hinz, 2011]{Carmona2011RiskNeutralMF}
Carmona, R.~A. and Hinz, J. (2011).
\newblock Risk-neutral models for emission allowance prices and option valuation.
\newblock {\em Manag. Sci.}, 57:1453--1468.

\bibitem[Carnero and Pérez, 2019]{CARNERO_leverage}
Carnero, M.~A. and Pérez, A. (2019).
\newblock Leverage effect in energy futures revisited.
\newblock {\em Energy Economics}, 82:237--252.
\newblock Replication in Energy Economics.

\bibitem[Chevallier, 2009]{CHEVALLIERRiskFactors}
Chevallier, J. (2009).
\newblock Carbon futures and macroeconomic risk factors: A view from the {EU} {ETS}.
\newblock {\em Energy Economics}, 31(4):614--625.

\bibitem[Chevallier and Sévi, 2011]{ChevallierRV}
Chevallier, J. and Sévi, B. (2011).
\newblock {On the realized volatility of the ECX {CO}$_2$ emissions 2008 futures contract: Distribution, dynamics and forecasting}.
\newblock {\em Annals of Finance}, 7(1):1--29.

\bibitem[Christoffersen et~al., 2009]{Christoffen2009}
Christoffersen, P., Heston, S., and Jacobs, K. (2009).
\newblock The shape and term structure of the index option smirk: Why multifactor stochastic volatility models work so well.
\newblock {\em Management Science}, 55:1914--1932.

\bibitem[Christoffersen et~al., 2013]{ChristoffenPricing}
Christoffersen, P., Heston, S., and Jacobs, K. (2013).
\newblock {Capturing Option Anomalies with a Variance-Dependent Pricing Kernel}.
\newblock {\em The Review of Financial Studies}, 26(8):1963--2006.

\bibitem[Cornago, 2022]{cornago2022eu}
Cornago, E. (2022).
\newblock The {EU} emissions trading system after the energy price spike.
\newblock {\em Centre for European Reform, Open Socienty European Policy Institute}.

\bibitem[Corsi, 2009]{CorsiRV}
Corsi, F. (2009).
\newblock {A Simple Approximate Long-Memory Model of Realized Volatility}.
\newblock {\em Journal of Financial Econometrics}, 7(2):174--196.

\bibitem[Corsi et~al., 2013]{CorsiFusVE}
Corsi, F., Fusari, N., and {La Vecchia}, D. (2013).
\newblock Realizing smiles: Options pricing with realized volatility.
\newblock {\em Journal of Financial Economics}, 107(2):284--304.

\bibitem[Corsi et~al., 2010]{CORSIJump}
Corsi, F., Pirino, D., and Renò, R. (2010).
\newblock Threshold bipower variation and the impact of jumps on volatility forecasting.
\newblock {\em Journal of Econometrics}, 159(2):276--288.

\bibitem[Corsi and Renò, 2012]{CorsiRenoJBES}
Corsi, F. and Renò, R. (2012).
\newblock Discrete-time volatility forecasting with persistent leverage effect and the link with continuous-time volatility modeling.
\newblock {\em Journal of Business and Economic Statistics}, 30(3):368--380.

\bibitem[Duffie et~al., 2000]{Duffie2000}
Duffie, D., Pan, J., and Singleton, K. (2000).
\newblock Transform analysis and asset pricing for affine jump-diffusions.
\newblock {\em Econometrica}, 68(6):1343--1376.

\bibitem[Fang et~al., 2024]{Fang}
Fang, M., Tan, K., and Wirjanto, T. (2024).
\newblock Valuation of carbon emission allowance options under an open trading phase.
\newblock {\em Energy Economics}, 131:107351.

\bibitem[Gerlagh et~al., 2020]{COVIDEUETS}
Gerlagh, R., Heijmans, R. J. R.~K., and Rosendahl, K.~E. (2020).
\newblock {COVID-19 Tests the Market Stability Reserve}.
\newblock {\em Environmental \& Resource Economics}, 76(4):855--865.

\bibitem[Gourieroux et~al., 1993]{Gourioeux1993}
Gourieroux, C., Monfort, A., and Renault, E. (1993).
\newblock Indirect inference.
\newblock {\em Journal of Applied Econometrics}, 8:S85--S118.

\bibitem[Gouriéroux and Monfort, 1997]{Gour1996}
Gouriéroux, C. and Monfort, A. (1997).
\newblock {\em {Simulation-based Econometric Methods}}.
\newblock Oxford University Press.

\bibitem[Hammoudeh et~al., 2014]{Hammoudeh}
Hammoudeh, S., Nguyen, D.~K., and Sousa, R.~M. (2014).
\newblock {Energy prices and {CO}$_2$ emission allowance prices: A quantile regression approach}.
\newblock {\em Energy Policy}, 70(C):201--206.

\bibitem[Heston, 1993]{Heston1993}
Heston, S.~L. (1993).
\newblock A closed-form solution for options with stochastic volatility with applications to bond and currency options.
\newblock {\em The Review of Financial Studies}, 6(2):327--343.

\bibitem[Hintermann, 2010]{HINTERMANN201043}
Hintermann, B. (2010).
\newblock Allowance price drivers in the first phase of the {EU} {ETS}.
\newblock {\em Journal of Environmental Economics and Management}, 59(1):43--56.

\bibitem[Hitzemann et~al., 2015]{HITZEMANN_intradayVol}
Hitzemann, S., Uhrig-Homburg, M., and Ehrhart, K.-M. (2015).
\newblock Emission permits and the announcement of realized emissions: Price impact, trading volume, and volatilities.
\newblock {\em Energy Economics}, 51:560--569.

\bibitem[Kim et~al., 2017]{KIM2017714}
Kim, J., Park, Y.~J., and Ryu, D. (2017).
\newblock Stochastic volatility of the futures prices of emission allowances: A {B}ayesian approach.
\newblock {\em Physica A: Statistical Mechanics and its Applications}, 465:714--724.

\bibitem[Majewski et~al., 2015]{Maje2015}
Majewski, A.~A., Bormetti, G., and Corsi, F. (2015).
\newblock {Smile from the past: A general option pricing framework with multiple volatility and leverage components}.
\newblock {\em Journal of Econometrics}, 187(2):521--531.

\bibitem[Mancini, 2009]{Mancini}
Mancini, C. (2009).
\newblock Non-parametric threshold estimation for models with stochastic diffusion coefficient and jumps.
\newblock {\em Scandinavian Journal of Statistics}, 36(2):270--296.

\bibitem[Mansanet-Bataller et~al., 2006]{Mansanet}
Mansanet-Bataller, M., Pardo, A., and Valor, E. (2006).
\newblock {CO}$_2$ prices, energy and weather.
\newblock {\em The Energy Journal}, 28:73--92.

\bibitem[Pacati et~al., 2018]{Heston++}
Pacati, C., Pompa, G., and Renò, R. (2018).
\newblock Smiling twice: The {H}eston++ model.
\newblock {\em Journal of Banking \& Finance}, 96:185--206.

\bibitem[Patton, 2011]{PattonQlike}
Patton, A.~J. (2011).
\newblock Volatility forecast comparison using imperfect volatility proxies.
\newblock {\em Journal of Econometrics}, 160(1):246--256.
\newblock Realized Volatility.

\bibitem[Renault, 1996]{Ren97}
Renault, E. (1996).
\newblock {Econometric Models of Option Pricing Errors}.
\newblock Technical report.

\bibitem[Rossi and de~Magistris, 2018]{RossiDemag}
Rossi, E. and de~Magistris, P.~S. (2018).
\newblock {Indirect inference with time series observed with error}.
\newblock {\em Journal of Applied Econometrics}, 33(6):874--897.

\bibitem[Rotfuß, 2009]{WaldeMarPriceFormationHF}
Rotfuß, W. (2009).
\newblock {Intraday price formation and volatility in the European Union emissions trading scheme: An introductory analysis}.
\newblock ZEW Discussion Papers 09-018, ZEW - Leibniz Centre for European Economic Research.

\bibitem[Yang et~al., 2016]{Yang}
Yang, S.~S., Huang, J.-W., and Chang, C.-C. (2016).
\newblock Detecting and modelling the jump risk of {CO}$_2$ emission allowances and their impact on the valuation of option on futures contracts.
\newblock {\em Quantitative Finance}, 16(5):749--762.

\bibitem[Zhang et~al., 2005]{zhang2s}
Zhang, L., Mykland, P., and Aït-Sahalia, Y. (2005).
\newblock A tale of two time scales.
\newblock {\em Journal of the American Statistical Association}, 100:1394--1411.

\end{thebibliography}
\end{document}